\documentclass{raa04}
\usepackage{graphicx,times}             
\usepackage{natbib}
\usepackage{amssymb,amsmath}
\bibpunct{(}{)}{;}{a}{}{,}

\usepackage[a4paper=true,dvipdfm=true,pagebackref=true]{hyperref}
\hypersetup{colorlinks = true, linkcolor = green, anchorcolor = red, citecolor = blue, filecolor = red, pagecolor = red, urlcolor = red}

\begin{document}

   \title{Three New Spiral Galaxies with Active Nuclei Producing Double Radio Lobes}

 \volnopage{Vol.0 (20xx) No.0, 000--000}      
   \setcounter{page}{1}          

   \author{X. Y. Gao
      \inst{1,2,3}
   \and Z. S. Yuan
      \inst{1,2}
   \and J. L. Han 
      \inst{1,2,3}
   \and Z. L. Wen
      \inst{1,2,3}
   \and S. S. Shan
      \inst{1,3}}

   \institute{National Astronomical Observatories, Chinese Academy of Sciences,
             Beijing 100101, China; xygao@nao.cas.cn; hjl@nao.cas.cn \\
             \and   
             CAS Key Laboratory of FAST, National Astronomical
             Observatories, Chinese Academy of Sciences, Beijing
             100101, China; \\
             \and
             School of Astronomy, University of Chinese Academy of
             Sciences, Beijing 100049, China \\
             \vs\no
   {\small Received~~20xx month day; accepted~~20xx~~month day}}

\abstract{Double radio lobes are generally believed to be produced by
  active nuclei of elliptical galaxies. However, several double-lobed
  radio sources have been solidly found to be associated with spiral
  galaxies. By cross-matching $\sim9\times10^5$ spiral galaxies
  selected from the Sloan Digital Sky Survey DR8 data with the full
  1.4~GHz radio source catalogs of NRAO VLA Sky Survey and Faint
  Images of Radio Sky at Twenty-centimeters, we identify three new
  spiral galaxies: J0326$-$0623, J1110+0321 and J1134+3046 that
  produce double radio lobes, and five double-lobed spirals previously
  known. By combining the newly discovered and all the other known
  cases in literature, we confirm the relation that more massive
  spirals could produce more powerful radio jets. We find that most of
  these spiral galaxies are located in a galaxy group or a poor
  cluster, in which the environment is denser than in the field, and
  about half of them are the central brightest galaxies in their
  parent system. We therefore suggest that the environment is one of
  the key factors for a spiral to produce double radio lobes.
  \keywords{galaxies: general --- galaxies: -- active --- galaxies:
    spiral --- radio continuum: galaxies --- galaxies: jets} }

    \authorrunning{X. Y. Gao et al.}  
    \titlerunning{Three spiral galaxies with double radio lobes} 
    \maketitle
%
%
\section{Introduction}           
\label{sect:intro}

Radio lobes powered by the supermassive black hole (SMBH) in the
centers of galaxies are usually hosted by elliptical galaxies. They
extend to several hundred kilo-parsecs which are much larger than
their optical counterparts. The point of view that double radio lobes
are exclusively hosted by elliptical galaxies has been challenged by
the discovery of the double-lobed radio source J0313$-$192
(J0315$-$1906), which was found to be hosted by a {\it spiral} galaxy
\citep{lok98,loyh01,kwol06}.

In the last two decades, a few more spiral galaxies hosting double
radio lobes have been revealed. \citet{hso+11} discovered the second
case, Speca (J1409$-$0302), which has episodic radio
emission. \citet{bvv+14} identified J2345$-$0449 with lobes extending
to an extraordinary length of $\sim$1.6~Mpc. \citet{mmm+16} reported
the serendipitous discovery of the double radio source of
MCG+07--47--10 (J2318+4314) with a low luminosity of $P_{\rm
  1.4GHz}\sim10^{22}~\rm W~Hz^{-1}$. Very recently, \citet{Vietri22}
identified a new source J0354$-$1340 with the double radio lobes
extending approximately 240~kpc.

Searches for spiral galaxies with double radio lobes have also been
made systematically by several groups.
\citet{mod+15} cross-matched the optical Galaxy Zoo ``superclean''
sample \citep{lss+08} with the Unified Radio Catalog of
\citet{Kimball08}, which includes radio sources from both the Faint
Images of the Radio Sky at Twenty-centimeters \citep[FIRST,][]{bwh95}
and the NRAO VLA Sky Survey \citep[NVSS,][]{ccg+98} data. They
reported a new spiral J1649+2635 with double radio lobes, which has a
radio power of about $\rm 10^{24}~W~Hz^{-1}$ at 1.4~GHz.
\citet{sis+15} cross-matched the FIRST catalog \citep{bwh95} with the
187\,005 spiral galaxies \citep{mvb15} from the Sloan Digital Sky
Survey \citep[SDSS,][]{yaa+00} Data Release 7 (DR7) to search for
coincidences of a core source within a radius of 3$\arcsec$ and also
the double lobes within a radius of 3$\arcmin$. They conducted an
additional search for the extended radio lobes with the NVSS data
\citep{ccg+98} for the obtained FIRST-SDSS matched objects and
identified four spiral galaxies with double radio lobes, among which
J1159+5820 \citep{kjz+12}, J1352+3126 \citep{Donzelli07}, and
J1649+2635 \citep{mod+15} were already reported previously, while
J0836+0532 was found for the first time.
\citet{Martinez16} collected 675\,874 spiral galaxies from several
spiral galaxy samples \citep[e.g.,][]{Huertas11,Willett13,ks16}, and
searched for the associated FIRST sources \citep{bwh95} within a
larger radius of 6$\arcmin$ under constraints of angular distances,
position angles and arm-length ratio of the double radio sources with
respect to the central optical galaxy. Concentrating on the extremely
symmetric and aligned radio lobes, they finally reported the
re-discovery of the known case of J1649+2635 \citep{mod+15}. Though
these efforts have been dedicated to searching for the spiral galaxies
hosting double radio lobes, only a handful of cases have been
confirmed to date.

It is also unclear why these spirals produce double radio lobes while
the vast majority of other spirals do not. Often radio lobes may be
triggered by the accretion of host galaxies from the over-dense
environments in their vicinity. For example, the source J0315$-$1906
is a member galaxy of the cluster A428 \citep{lok98}. J1409$-$0302 and
J1649+2635 are the brightest galaxies of their parent systems
\citep{hso+11,mod+15}, and J2318+4314 is located close to the galaxy
groups NGC~7618 and UGC~12491 \citep{mmm+16}. However, \citet{sis+15}
found that J0836+0532 and J1352+3126 are in galaxy groups with very
limited members and listed them as field galaxies together with
J1159+5820 (see their Table~7). Therefore, a large sample of such
galaxies is needed to investigate the environmental effect on radio
lobes. \citet{Wu22} recently analyzed the optical images from the
Hubble Space Telescope of a sample of galaxies with extended double
radio lobes seen from FIRST \citep{bwh95}, and found that 18 disk
galaxies have high probability of genuine association. Some of these
disk galaxies have a small inclination angle and show clear spiral
patterns.

We notice that \citet{ks16} classified the broad morphological types
of $\sim3\times10^6$ galaxies in the SDSS \citep{yaa+00} Data Release
8 (DR8) by analyzing images of galaxies with computer programs, and
their pipeline picked out $\sim9\times10^5$ spiral galaxies. Here, we
take this large sample as the optical basis of spiral galaxies, and
cross-match them with the full radio source catalogs of NVSS
\citep{ccg+98} and FIRST \citep{bwh95}. We discover three new spiral
galaxies with double radio lobes. The paper is organized as
follows. In Sect.~2, we introduce the data sets used to identify the
double-lobed spiral galaxies, and also the procedure of
identification. We show the results and discuss the properties of the
galaxies and their environments in Sect.~3. The concluding remarks are
given in Sect.~4.

Throughout the paper, we adopt a flat $\Lambda$CDM cosmology with
$H_{0} = 70~km\ s^{-1}\ Mpc^{-1}$, $\Omega_{m}$ = 0.3 and
$\Omega_{\Lambda}$ = 0.7.

\section{Data and Search Strategy}
\label{sect:Obs}

\subsection{Data}
Via an automatic computer program, \citet{ks16} classified
approximately 9$\times10^5$ spiral galaxies out of $\sim$3 $\times$
10$^{6}$ galaxies observed in the SDSS \citep{yaa+00} DR8. The
classifications for spiral galaxies and elliptical galaxies show good
agreement with those of the Galaxy Zoo debiased ``superclean'' sample
\citep{lss+08} and the agreement rate is claimed to reach 98\% when
the ``classification certainty'' $p \ge 0.54$.

Two catalogs were released by \citet{ks16}. The
``catalog.dat''\footnote{https://cdsarc.cds.unistra.fr/viz-bin/cat/J/ApJS/223/20$\#$/browse}
is the catalog of the broad morphology of SDSS galaxies with only the
classification certainty $p$ indicated, and the ``spec.dat'' including
both indications of ``$p$'' and morphological remarks (Elliptical,
Spiral, Star) is the morphological catalog of the SDSS objects with
spectra. In this work, we first take all the spiral galaxies in the
two catalogs with $p \ge 0.54$. However, we notice that some known
spirals such as J0836+0532 and J1649+2635, as shown in
Table~\ref{sdragn}, have a classification certainty of $p$ = 0.23 and
$p$ = 0.37, respectively, less than 0.54. In order not to miss many
real spirals, we simply included all galaxies marked as ``Spiral'' in
the ``spec.dat'' of \citet{ks16}, regardless of the values of $p$.
Therefore, all 366\,836 entries in the ``spec.dat'' marked as
``Spiral'' and 1\,184\,922 entries with $p \ge 0.54$ in
``catalog.dat'' are used to search for associated double radio
lobes. The duplicates in the two catalogs are treated at the final
stage when inspecting the association between optical and radio
images.

\begin{table*}[t]
\centering
\tabcolsep 1.5pt
\caption{Parameters for 17 spiral galaxies that host double radio lobes.} 
\label{sdragn}
\footnotesize
\begin{tabular}{lcrrccrccccccc}
\hline\hline
\multicolumn{1}{c}{Name}& Ref.1 & \multicolumn{1}{c}{R.A.}  &\multicolumn{1}{c}{Decl.}  &\multicolumn{1}{c}{$z$}  &\multicolumn{1}{c}{$S_{\rm tot}$}  &\multicolumn{1}{c}{$P_{\rm 1.4~GHz}$}  &\multicolumn{1}{c}{$p$}    &\multicolumn{1}{c}{${\rm log}_{10}M_{\ast}$}  &\multicolumn{1}{c}{Environment} &\multicolumn{1}{c}{$N_{\rm gal}$} &\multicolumn{1}{c}{Charc.}  & Ref.2  \\
\multicolumn{1}{c}{} &  &\multicolumn{1}{c}{(J2000)}       &\multicolumn{1}{c}{(J2000)}   & &\multicolumn{1}{c}{(mJy)}   &\multicolumn{1}{c}{({\rm W~Hz$^{-1}$)}}    &  &\multicolumn{1}{c}{($M_{\odot}$)} &\multicolumn{1}{c}{} &\multicolumn{1}{c}{}   &\multicolumn{1}{c}{}  &\multicolumn{1}{c}{} \\
\multicolumn{1}{c}{(1)}   &\multicolumn{1}{c}{(2)}       &\multicolumn{1}{c}{(3)}   &\multicolumn{1}{c}{(4)}  &\multicolumn{1}{c}{(5)}     &\multicolumn{1}{c}{(6)}  &\multicolumn{1}{c}{(7)}  &\multicolumn{1}{c}{(8)} &\multicolumn{1}{c}{(9)} &\multicolumn{1}{c}{(10)} &\multicolumn{1}{c}{(11)} &\multicolumn{1}{c}{(12)} &\multicolumn{1}{c}{(13)}\\
\hline
J0209+0750   & [1]            & 32.26979   & 7.83458    & 0.251  & 394  & $7.58\times10^{25}$   &0.37      & 11.45      & field    &  0       & $\cdots$ & $\cdots$ \\
J0315$-$1906 & [2]            &  48.96708  &$-$19.11233 & 0.067  & 100  & $1.05\times10^{24}$   & $\cdots$ & 10.88      & cluster  & 11       & member   & [2] \\
J0354$-$1340 & [3, 4]         &  58.63688  &$-$13.66868 & 0.076  &  15  & $2.05\times10^{23}$   & $\cdots$ & 11.13      & $\cdots$ & $\cdots$ & $\cdots$ & $\cdots$ \\
J0806+0624   & [1]            & 121.74358  & 6.41483    & 0.112  & 5    & $1.67\times10^{23}$   & $\cdots$ & 10.19      & field    &  1       & $\cdots$ & $\cdots$ \\
J0836$+$0532 & [5]            & 129.23278  &  5.54502   & 0.099  &  62  & $1.50\times10^{24}$   &0.23      & 10.91      & group    &  4       & BGG      & [5] \\ 
J1328+5710   & [1]            & 202.03879  & 57.17314   & 0.032  & 22   & $4.91\times10^{22}$   & $\cdots$ & 8.76       & field    &  0       & $\cdots$ & $\cdots$ \\
J1656+6407   & [1]            & 254.08583  & 64.13136   & 0.212  & 75   & $9.76\times10^{24}$   &0.91      & 11.50      & field    &  5       & $\cdots$ & $\cdots$ \\
J2318$+$4314 & [6]            & 349.63650  & 43.24692   & 0.012  &  17  & $5.26\times10^{21}$   & $\cdots$ &  9.12      & group    & $\cdots$ & member   & [6] \\ 
J2345$-$0449 & [7]            & 356.38625  &$-$4.82372  & 0.076  & 181  & $2.47\times10^{24}$   & $\cdots$ & 11.20      & group    &  7       & BGG      & [13] \\ 
\hline
J1128$+$2417 & [1, 0]         & 172.04848  & 24.29636   & 0.169$^{b}$  &  69  & $5.34\times10^{24}$   &0.66     & 10.57   & group   &  4    & member  & [0] \\ 
J1159$+$5820 & [8, 5, 0]      & 179.77361  & 58.34330   & 0.054  & 338  & $2.25\times10^{24}$   &0.85      & 10.88      & group    &  1    & BGG     & [14] \\ 
J1352$+$3126 & [9, 5, 0]      & 208.07450  & 31.44625   & 0.045  &4844  & $2.21\times10^{25}$   &0.98      & 10.91      & group    &  1    & BGG     & [15] \\ 
J1409$-$0302 & [10, 0]        & 212.45355  &$-$3.04237  & 0.138  & 139  & $6.91\times10^{24}$   &0.67$^s$  & 11.14      & cluster  & 10    & BCG     & [15] \\ 
J1649$+$2635 & [11, 5, 12, 0] & 252.35005  & 26.58405   & 0.055  & 157  & $1.09\times10^{24}$   &0.37$^s$  & 10.88      & cluster  & 12    & BCG     & [14] \\ 
\hline
J0326$-$0623 & [0]            &  51.59929  &$-$6.38431  & 0.180  & 6$^{\diamond}$  &   $4.94\times10^{23~\diamond}$   &0.67     & 10.87  & cluster & 13    & BCG     & [16] \\ 
J1110$+$0321 & [0]            & 167.60458  &  3.36078   & 0.030  & 587  & $1.17\times10^{24}$   &0.95      &  9.14      & group    & 7     & member  & [15] \\
J1134$+$3046 & [0]            & 173.58466  & 30.77959   & 0.046  & 380  & $1.82\times10^{24}$   &0.94      &  9.51      & group    & 4     & member  & [15] \\
\hline 
\end{tabular}  
\tablecomments{0.9\textwidth}{
Columns (1) -- (2): source name and the 
reference to find double lobes; Here Ref.1 is indicated by numbers:
[0]= this work; [1] = \citet{Wu22}; [2] = \citet{lok98}; [3] = \citet{Chen20}; 
[4] = \citet{Vietri22}; [5] = \citet{sis+15}; [6] = \citet{mmm+16}; [7] = \citet{bvv+14}; 
[8] = \citet{kjz+12}; [9] = \citet{Donzelli07}; [10] = \citet{hso+11}; 
[11] = \citet{mod+15}; [12] = \citet{Martinez16};
Columns (3) -- (5): R.A., decl.,  
and redshift of spiral galaxies respectively. The redshift information
is taken from NED (https://ned.ipac.caltech.edu)
except for J0209+0750, J0806+0624, J1328+5710 and J1656+6407, 
which are from reference [1] and J1128$+$2417 labeled with ``$b$'', 
which is the photo-z from SDSS DR8;
Column (6): total radio continuum flux density measured by the NVSS
at 1.4~GHz, ``$\diamond$'' indicates that the flux density is uncertain due 
to the mixture of an unrelated source (see Sect.~\ref{Notes}).
Column (7): radio powers calculated according to Equation~\ref{power}. 
Column (8): classification certainty taken from the ``catalog.dat'' of \citet{ks16} and the 
marker ``$s$'' indicates the classification certainty from the ``spec.dat'' of \citet{ks16}; 
Column (9): stellar mass of the galaxy;
Columns (10) -- (12) : environment as being a group or a cluster, the number of galaxies in the group/cluster and 
the character of the spiral in the environment, as being the BCG/BGG or just a member, respectively.
%
Column (13): reference indicating spirals are in galaxy groups/clusters: [0, 2, 5, and 6] are the same as in Column (2), 
together with [13] = \citet{svc+16}; [14] = \citet{Tempel18}; [15] = \citet{Tempel12}; [16] = \citet{Yang07}.
}
\end{table*}

The radio counterparts of the optical spiral galaxies are searched in
the NVSS \citep{ccg+98} and FIRST \citep{bwh95} source catalogs. The
NVSS was carried out with the Very Large Array (VLA) - D configuration
at the frequency of 1.4~GHz with an angular resolution of
$\sim45\arcsec$. The survey covers the entire sky north of
decl. $\delta = -40\degr$. Over 1.8~million discrete sources brighter
than $\sim$2.5~mJy (5$\sigma$ level) were compiled into the associated
catalog\footnote{https://www.cv.nrao.edu/nvss/NVSSlist.shtml}. The
positional accuracy of NVSS is about 1$\arcsec$ for strong sources and
7$\arcsec$ for faint sources. The FIRST observations were conducted at
the same frequency, but with a much better angular resolution of
$\sim5\arcsec$ by using the VLA - B configuration array. It has a
sensitivity of about 0.13~mJy beam$^{-1}$. The sky coverage of FIRST
is limited within the northern and southern Galactic cap regions of
about 10\,000 square degrees in total, which is less than one third
that of NVSS. The FIRST catalog contains over $9.4\times10^{5}$
entries. For the sources whose flux density is higher than 1~mJy, the
radius of the 90\% positional confidence error circle is less than
1$\arcsec$. The latest FIRST source catalog of Version
14Dec17\footnote{http://sundog.stsci.edu/first/catalogs/readme.html}
was used in this study.

\subsection{Search Strategy}
Based on the experiences gained from previous work
\cite[e.g.][]{yhw16}, we learn that the 5$\arcsec$ angular resolution
of FIRST could resolve out extended radio structures so that diffuse
radio lobes can be missed, such as the case for J1409--0302
\citep[Speca,][]{hso+11}. Therefore we take the NVSS data as the
fundamental basis and the FIRST data are used as an auxiliary database
for radio-lobe identification. The radius to search for the radio
counterpart of the central optical spiral galaxy is tricky: the larger
the radius is set, the more radio sources around the central galaxy
one would get, and then the more time would be consumed for
distinguishing the true association. Therefore a trade-off should be
made on setting the searching area to balance the time consumption. By
reviewing the known cases to date \citep[e.g.][]{lok98, hso+11,
  bvv+14, mod+15, sis+15, mmm+16}, we search for the NVSS and FIRST
sources within 800~kpc around the central optical spirals if the
redshift information is available; otherwise a radius of 3.5$\arcmin$
around the spiral is set for the NVSS sources and 30$\arcsec$ for the
FIRST sources if the redshift is unknown.

Spiral galaxies with double radio lobes may have various appearances
in the radio images of NVSS and FIRST. In the low resolution image of
the NVSS, they could have (1) a central core with distinct double
radio lobes, e.g. J1352+3126 \citep[see Fig.~4 in ][]{sis+15}; (2)
unresolved central core and double radio lobes, i.e. showing a
structure that is elongated, such as J1649+2635 \citep[see Fig.~5 in
][]{sis+15}; and (3) distinct double-lobe structure without a core
component intrinsically or extrinsically. In the high resolution image
of FIRST, the spirals that host double radio lobes could show (1) a
central core with distinct double radio lobes, e.g. J1649+2635
\citep{sis+15}; (2) only a central core, because the extended lobes
are resolved out by the small synthesis beam; and (3) distinct
double-lobe structures without a core, e.g. J1409$-$0302
\citep{hso+11}. The real cases can be any reasonable combinations of
the above possibilities for the NVSS and FIRST data in their common
surveyed area. However, such loose constraints will yield too many
output images, which are very difficult to be checked
manually. According to the observational fact that the associated
radio lobes are generally among the closest sources to the optical
center, we therefore only consider the four closest radio sources to
the central galaxy for association. Unlike \citet{Martinez16} who
searched for extremely symmetric and collimated jets, we allow the
angle between the two radio lobes (any pair of the four closest radio
sources) to vary in the range of $180\pm20\degr$ with respect to the
central galaxy. To avoid missing the cases of blended core and lobes,
we also accept the cases which have one or more radio sources close
enough ($\leqslant$ 22.5$\arcsec$, half of the beam size of the NVSS)
to the central spirals. Finally, the quantity of the images that
qualified for the above conditions is largely reduced and become
suitable for eye inspections. About 200\,000 images in total are left
and inspected manually.

The probable candidates are then picked out and further examined in
composite images which combine both information from radio and
optical. The optical images are taken from the Dark Energy
Spectroscopic Instrument (DESI) Legacy Imaging
Survey\footnote{http://legacysurvey.org/} \citep{dsl+19}, which is
deeper and has a better quality than the SDSS. More importantly, the
galaxy images observed by DESI can be well modeled by ``{\it The
  Tractor}'' with the point spread function considered \citep[see][for
  details]{dsl+19}. The model-subtracted image, namely, the residual
image, can be conveniently used to determine the existence of spiral
patterns in galaxies.

As in previous discoveries introduced in Sect.~\ref{sect:intro}, the
identified host galaxies with double radio lobes all show spiral
patterns. Even for J0315$-$1906, which is somehow edge-on,
\citet{lok98} claimed the detection of a spiral structure through a
deep $B$-band exposure. We follow the same discipline in this work
that a spiral pattern must be visible for the central optical
galaxy. Edge-on galaxies, which appear as disks are therefore not
considered.

\section{Result and Discussion}
\label{sect:result}

\begin{figure*}[!thp]
\centering
\includegraphics[angle=0,width=0.39\textwidth]{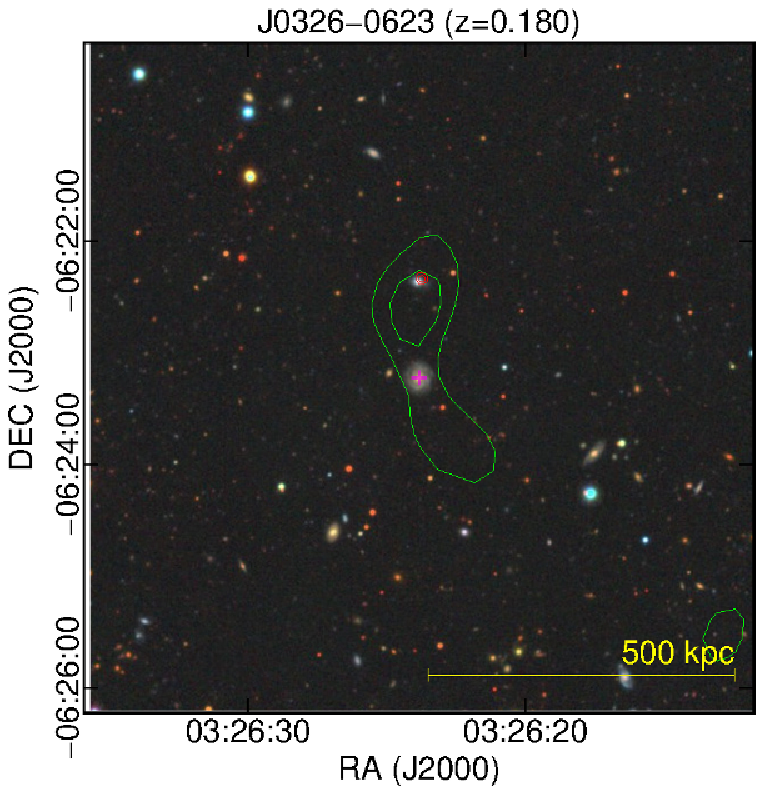}
\hspace{8mm}
\includegraphics[angle=0,width=0.39\textwidth]{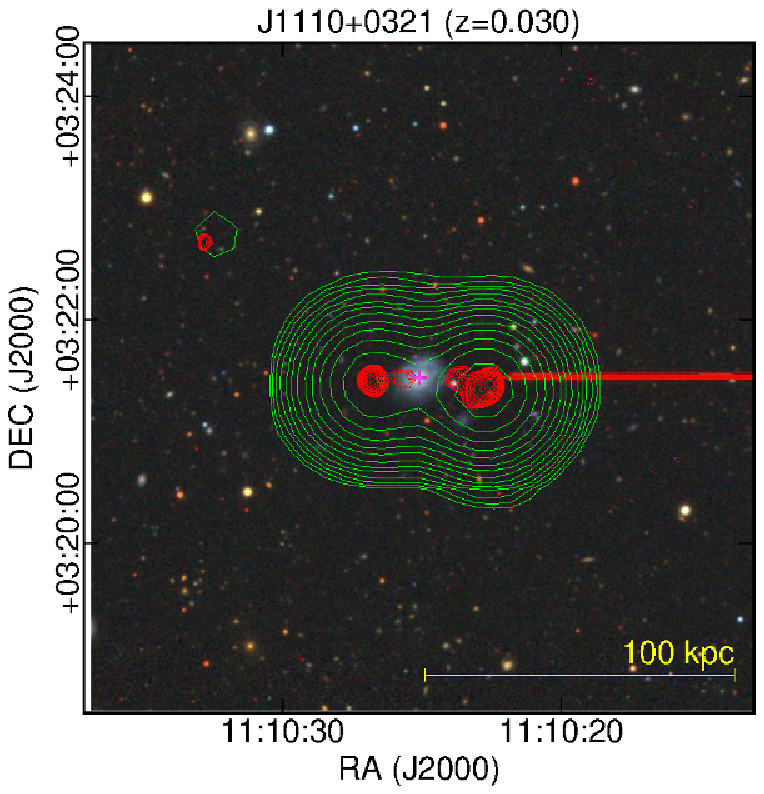}\\
\hspace{7mm}
\includegraphics[angle=0,width=0.174\textwidth]{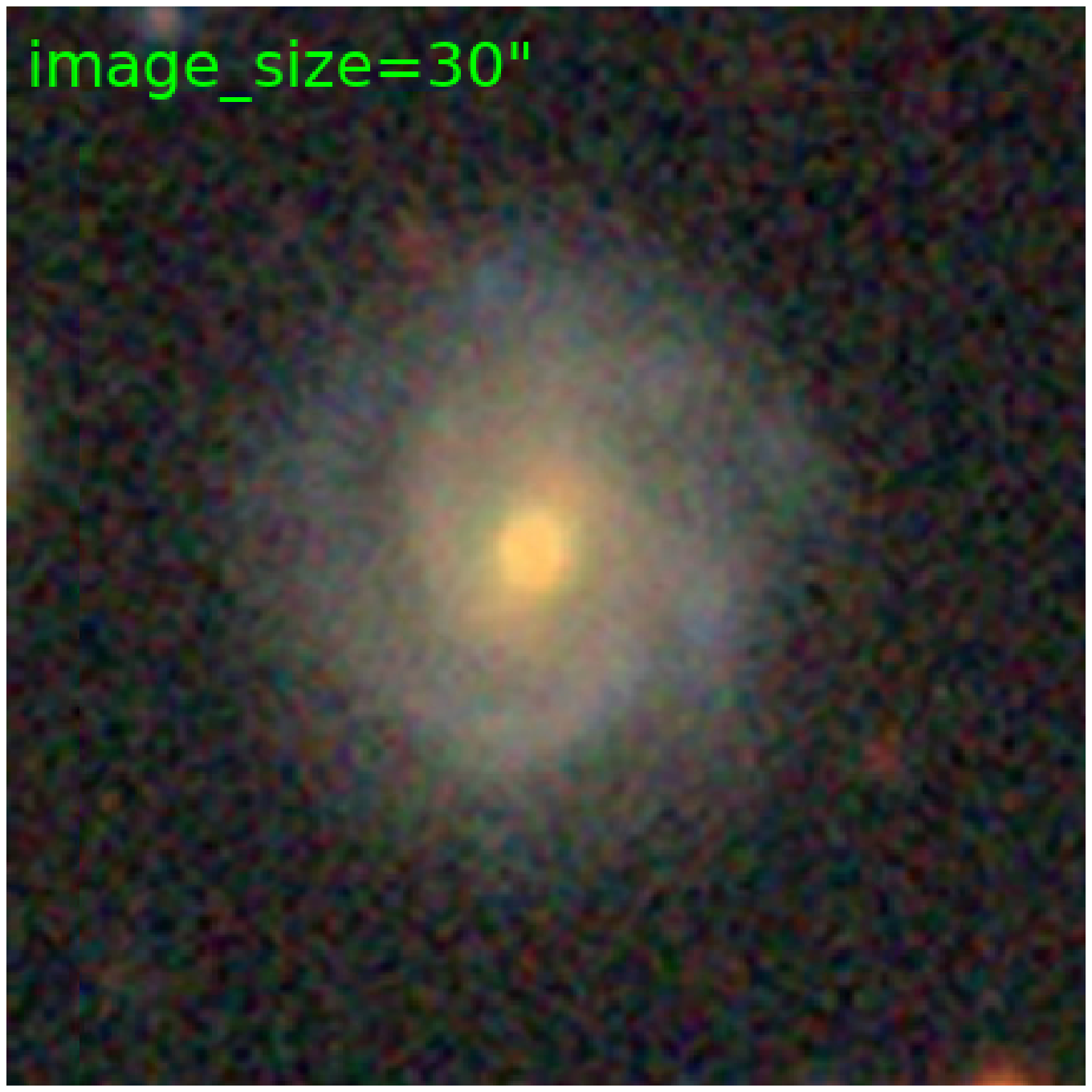}
\includegraphics[angle=0,width=0.174\textwidth]{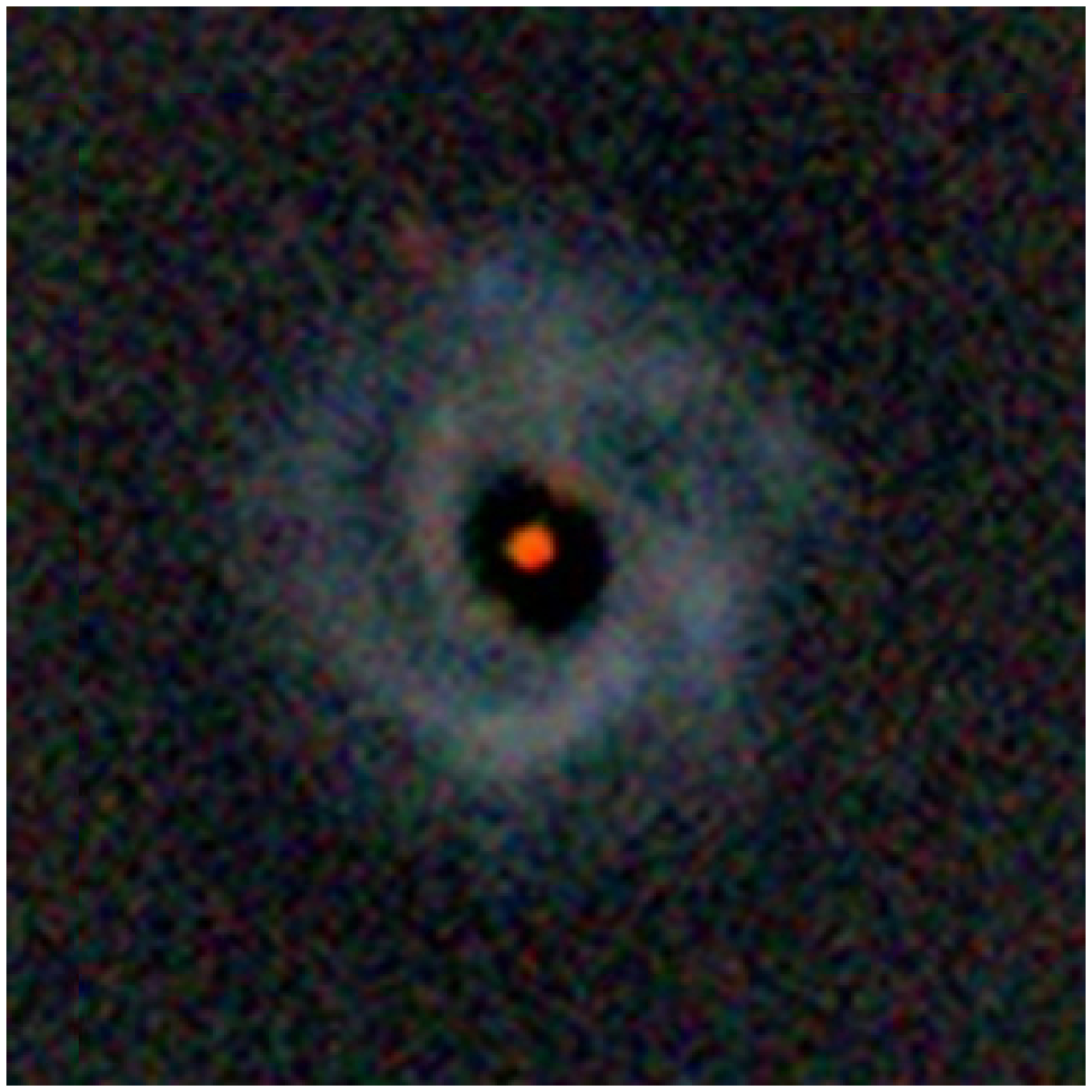}
\hspace{1.4cm}
\includegraphics[angle=0,width=0.174\textwidth]{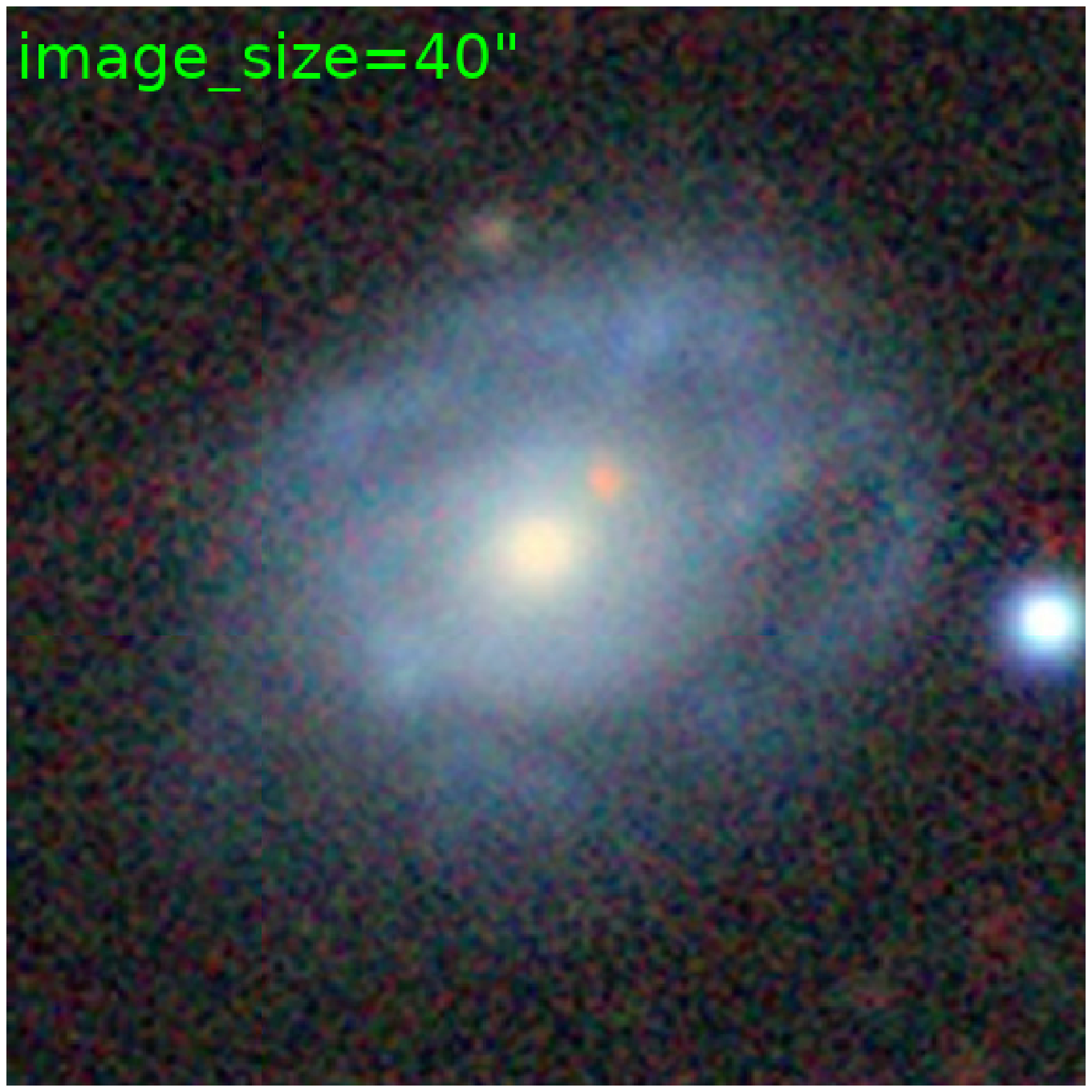}
\includegraphics[angle=0,width=0.174\textwidth]{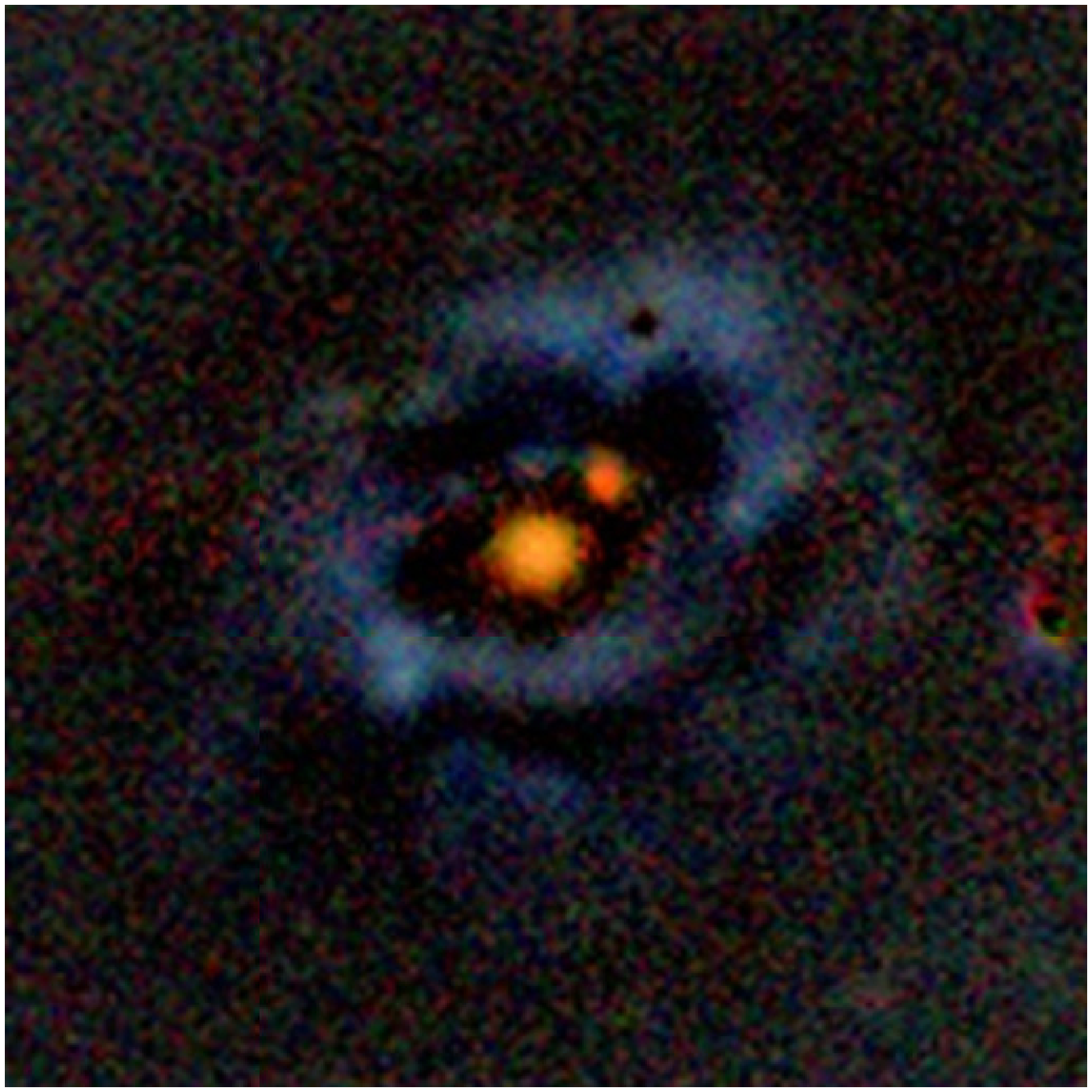}\\[0.5cm]
\includegraphics[angle=0,width=0.39\textwidth]{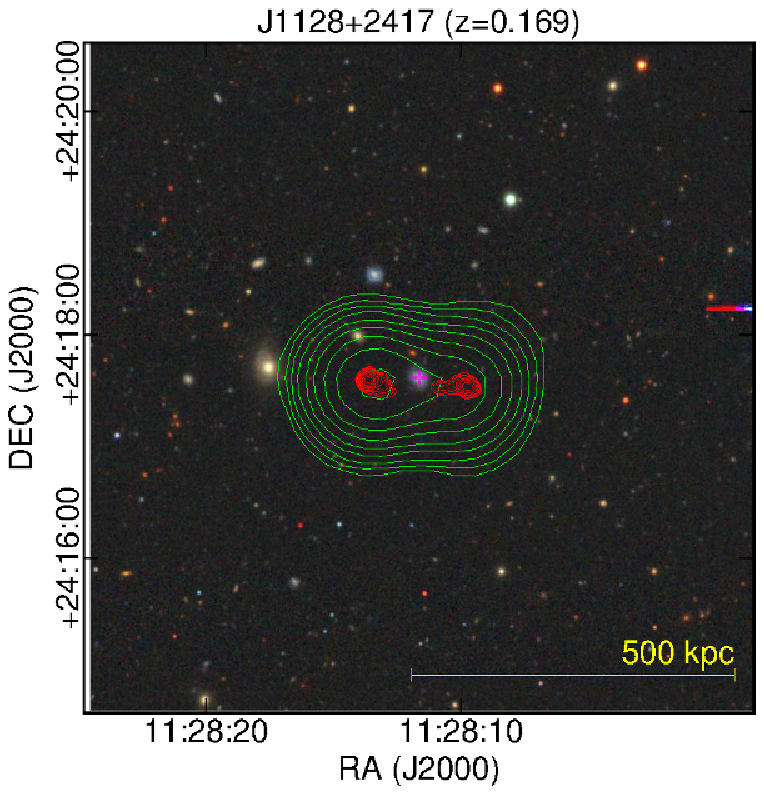}
\hspace{8mm}
\includegraphics[angle=0,width=0.39\textwidth]{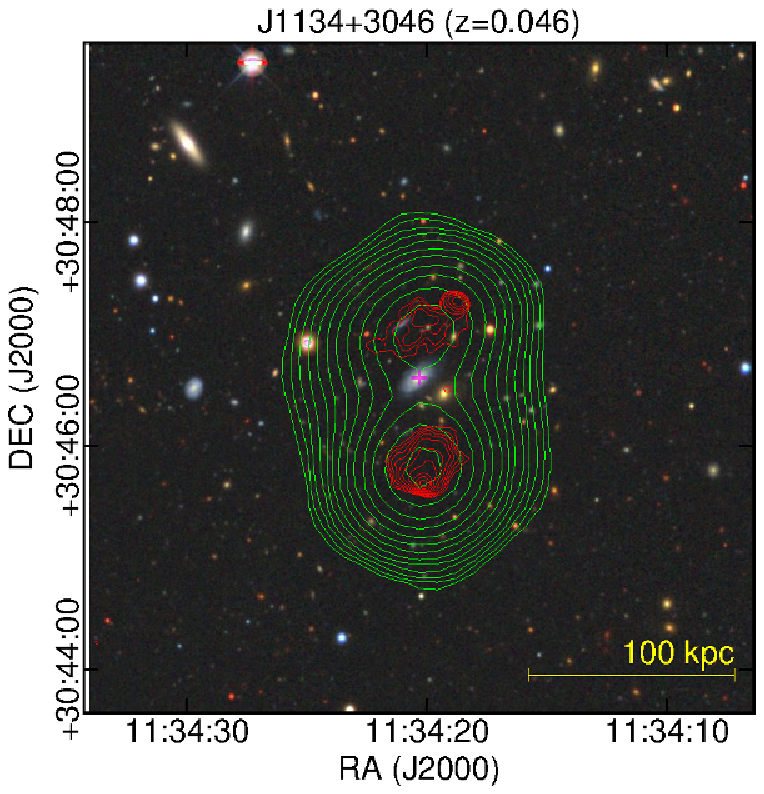}\\
\hspace{7mm}
\includegraphics[angle=0,width=0.174\textwidth]{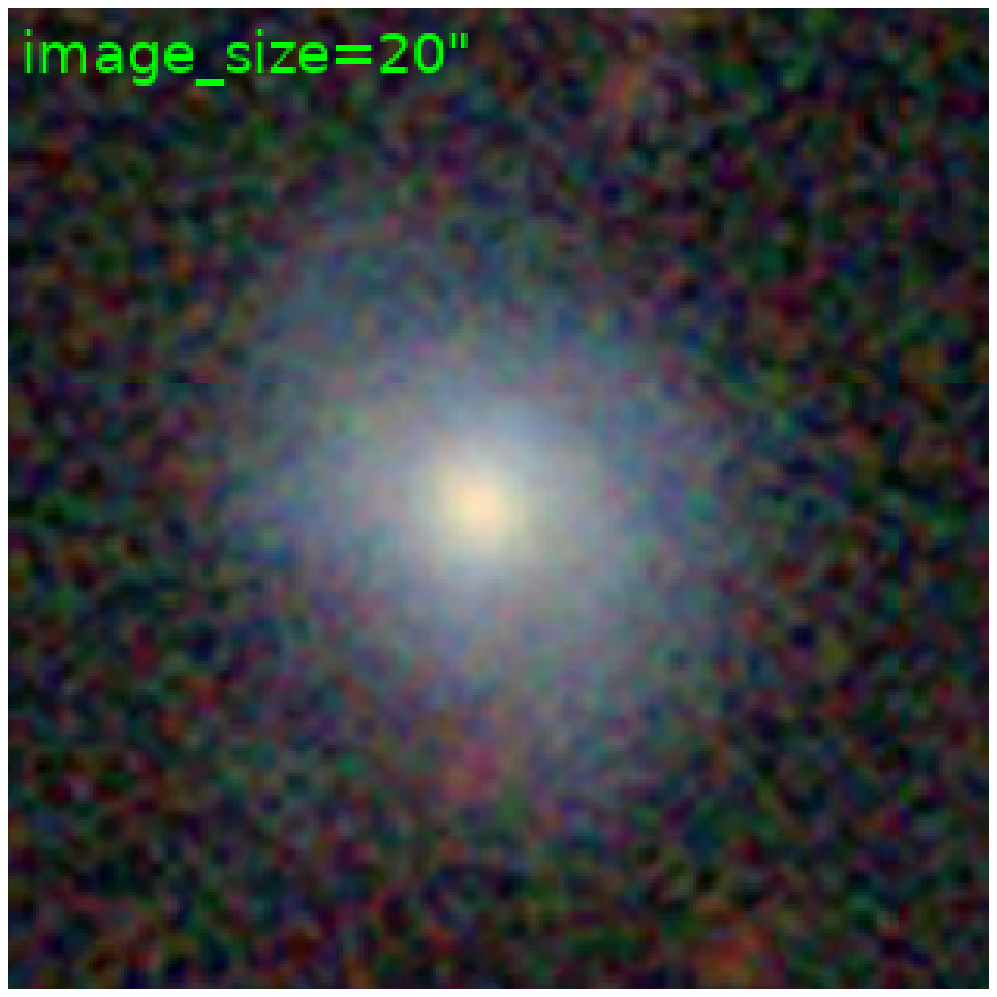}
\includegraphics[angle=0,width=0.174\textwidth]{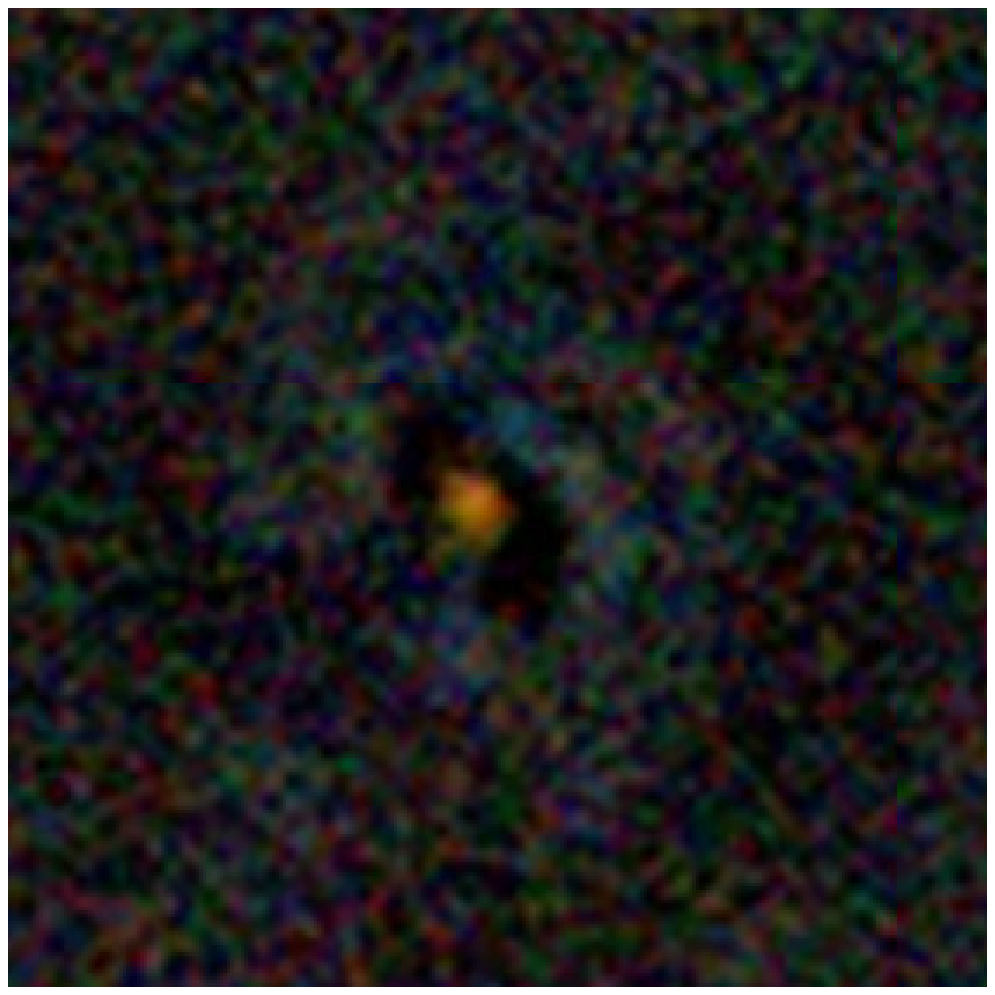}
\hspace{1.4cm}
\includegraphics[angle=0,width=0.174\textwidth]{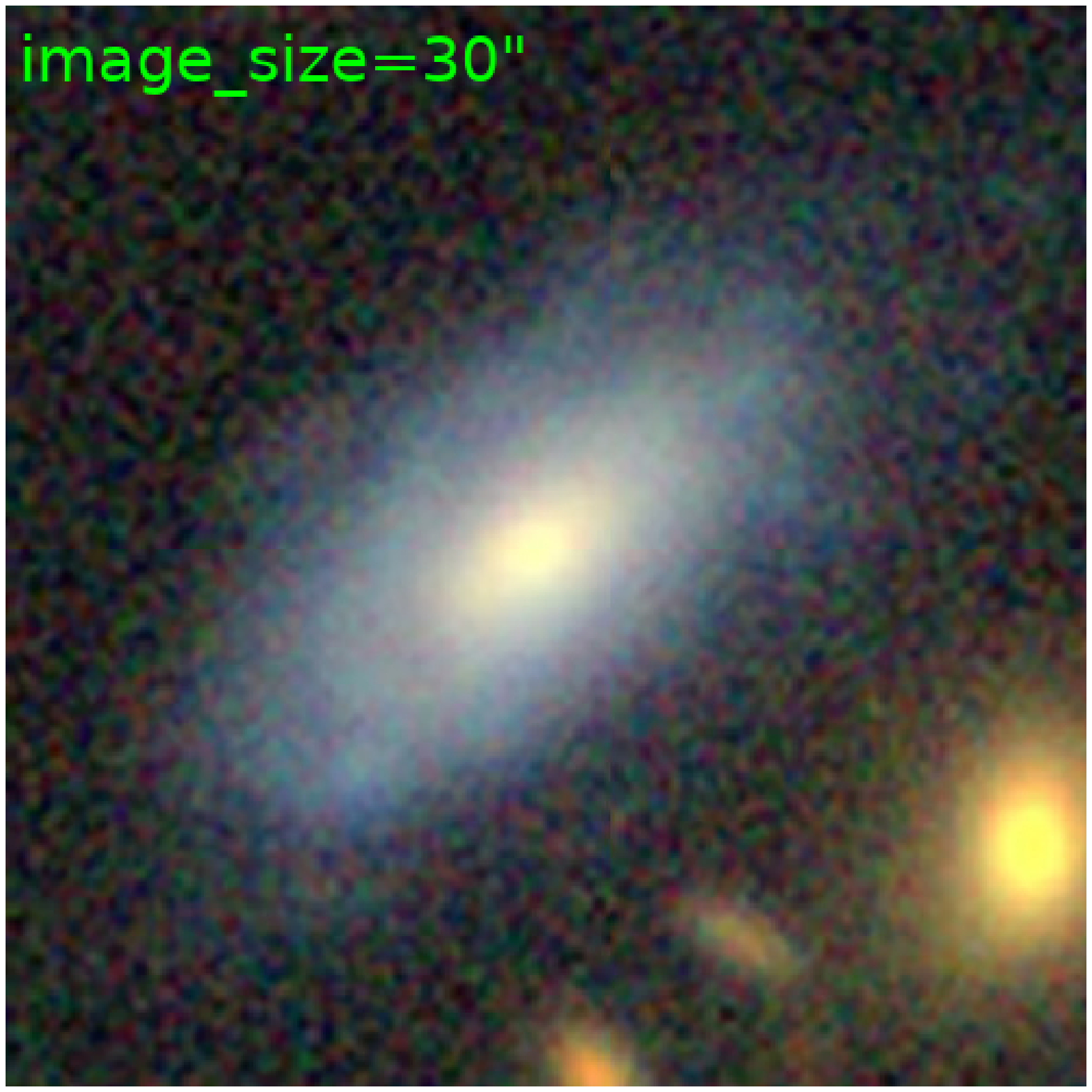}
\includegraphics[angle=0,width=0.174\textwidth]{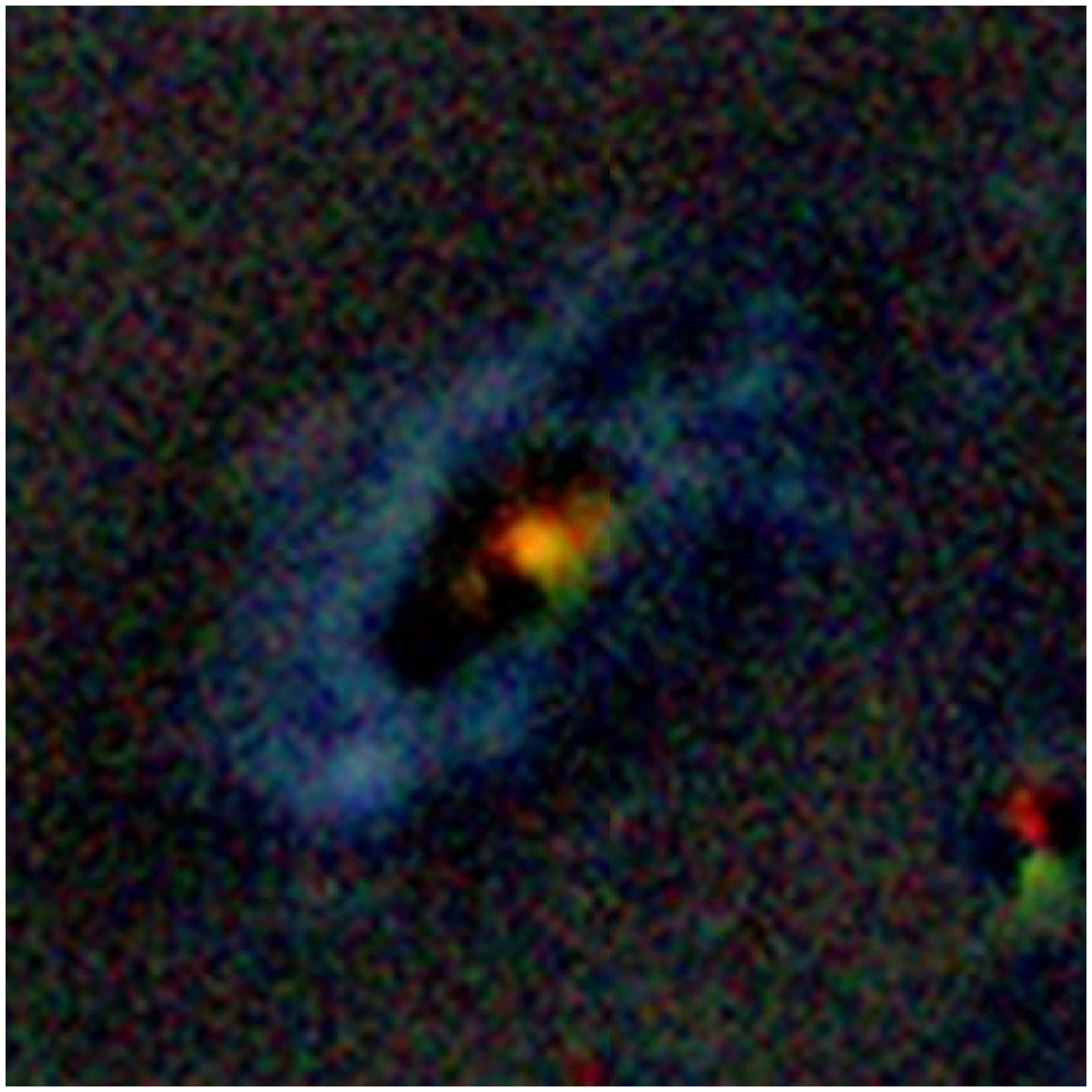}\\
\caption{Images for the eight spiral galaxies hosting double radio
  lobes identified in this work. The panels with large figures show
  the optical DESI images \citep{dsl+19} overlaid with radio contours,
  green for NVSS \citep{ccg+98} and red for FIRST \citep{bwh95}. Both
  the NVSS and FIRST contours satisfy $\langle S_{\rm
    bg}\rangle+5\times2^{\rm n/2}\sigma$~mJy beam$^{-1}$, here $n=0,
  1, 2, ...$. The source name and redshift are labeled on top of each
  plot. The cross indicates the center of the radio images. The
  physical scale is shown at the bottom-right corner. The panels with
  small figures are the zoomed-in DESI images and the model-subtracted
  residual images \citep{dsl+19} for the eight spiral galaxies, with
  the image size marked at the top-left corner.}
\end{figure*}

\addtocounter{figure}{-1}
\begin{figure*}[!thp]
\centering
\includegraphics[angle=0,width=0.39\textwidth]{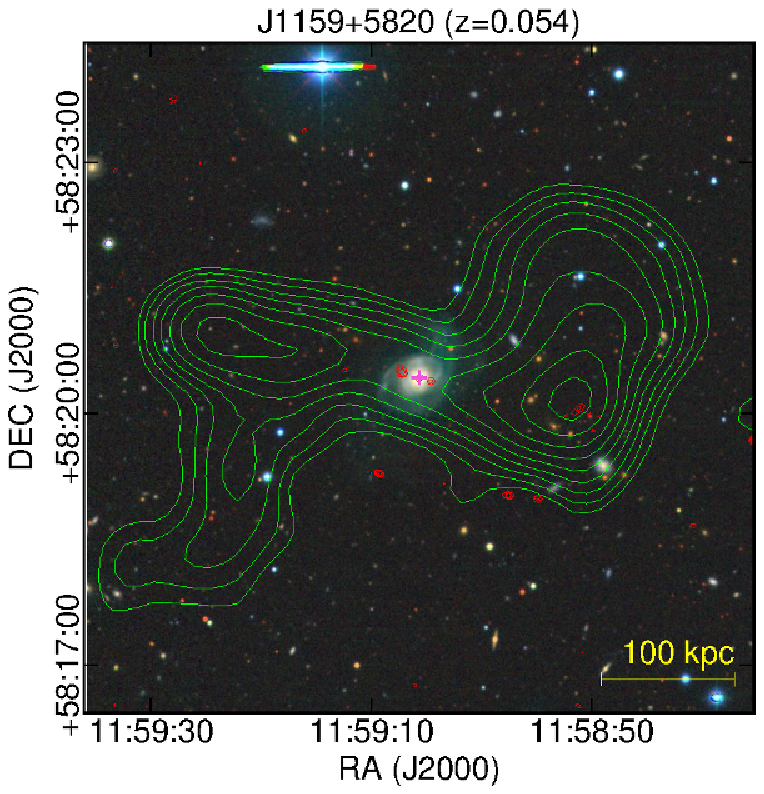}
\hspace{8mm}
\includegraphics[angle=0,width=0.39\textwidth]{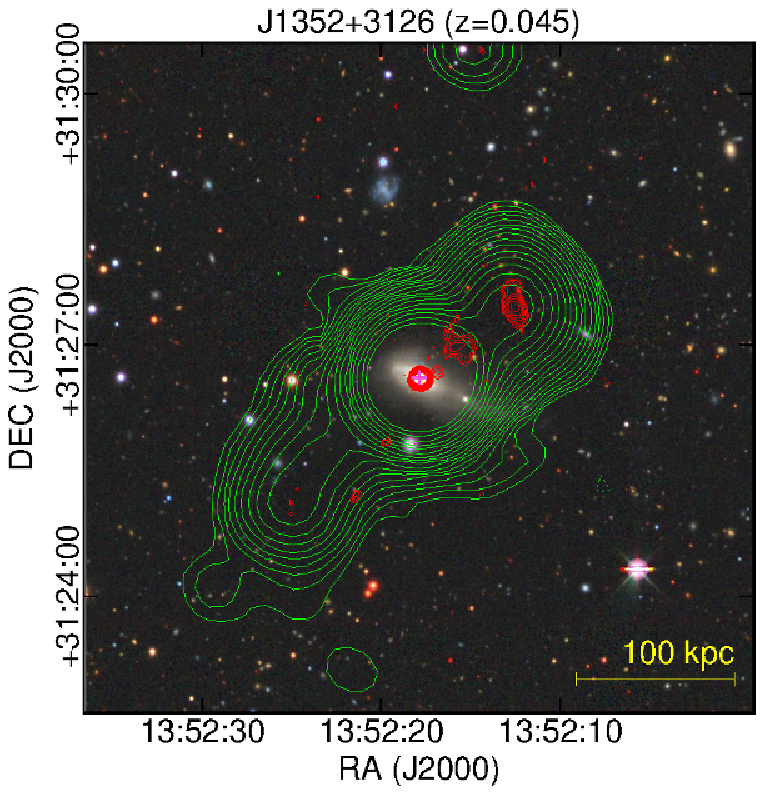}\\
\hspace{7mm}
\includegraphics[angle=0,width=0.174\textwidth]{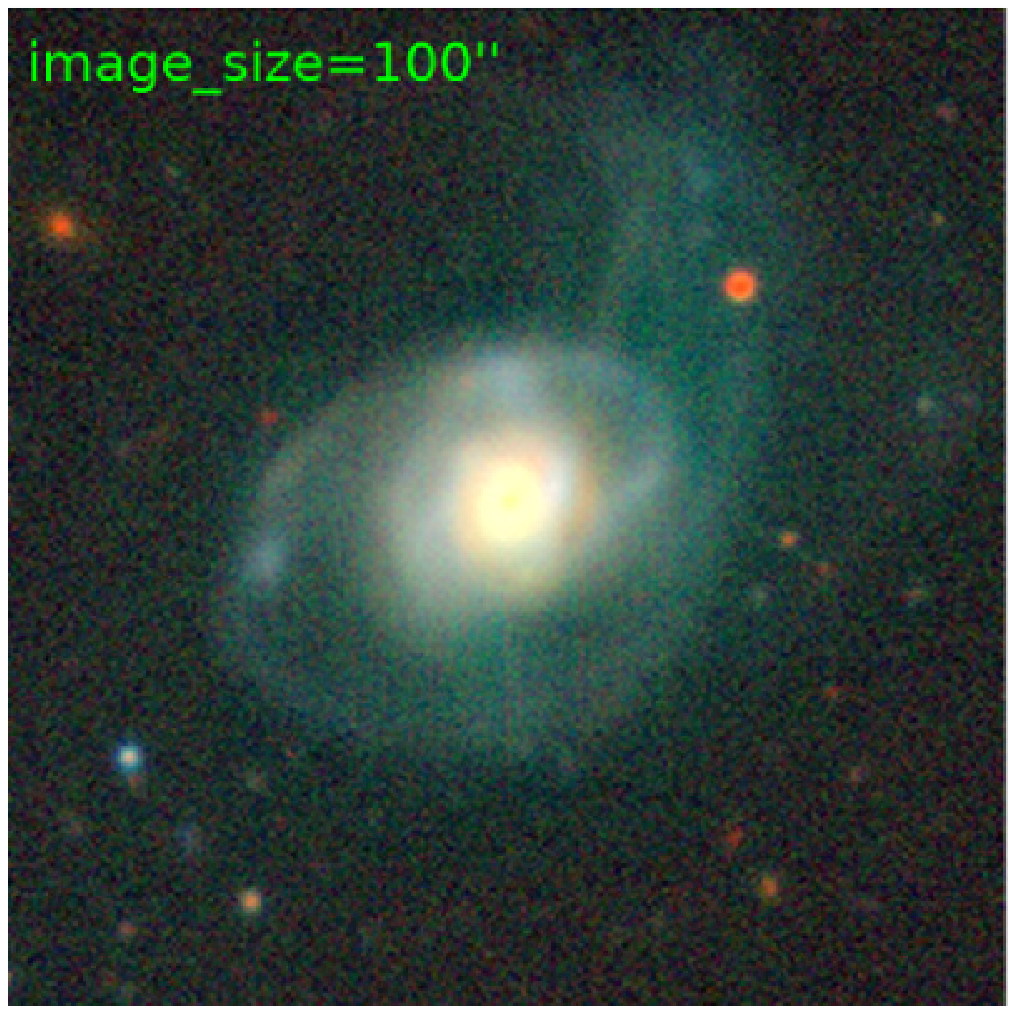}
\includegraphics[angle=0,width=0.174\textwidth]{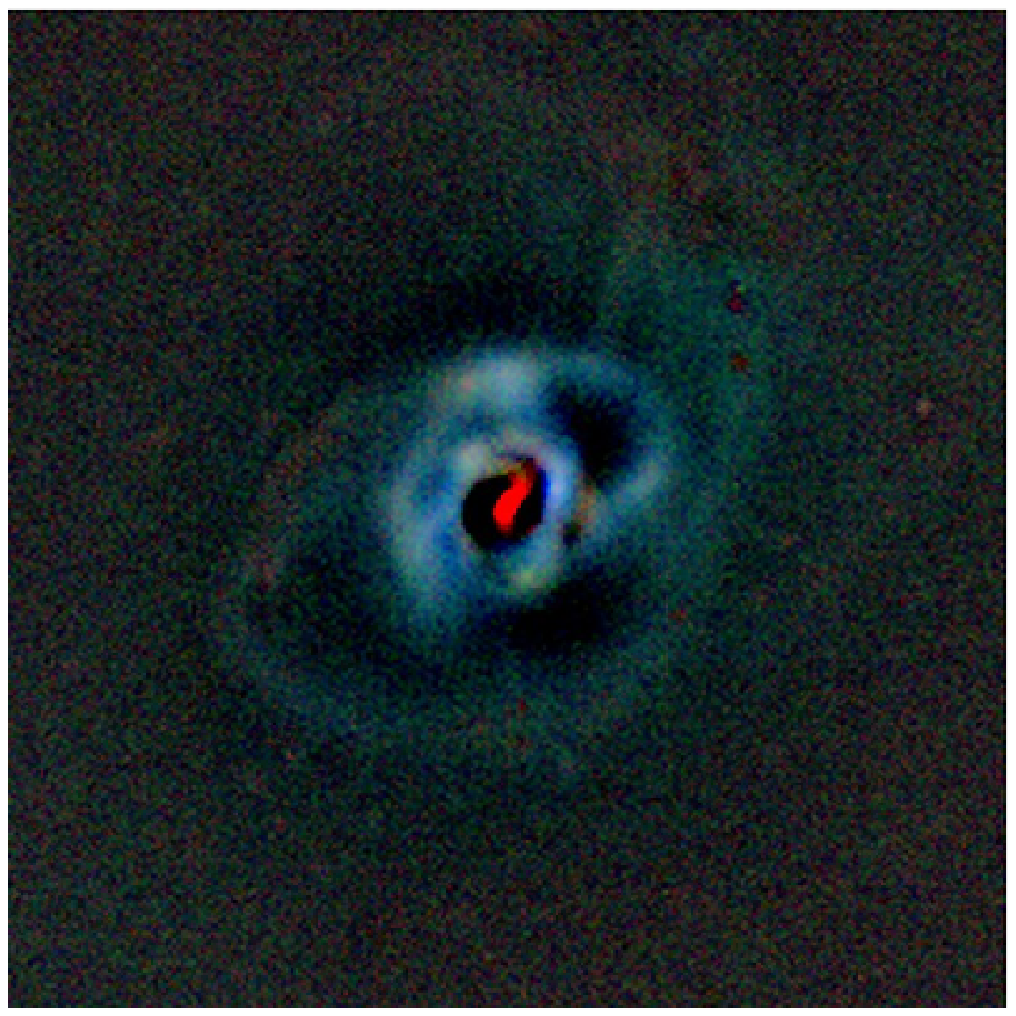}
\hspace{1.4cm}
\includegraphics[angle=0,width=0.174\textwidth]{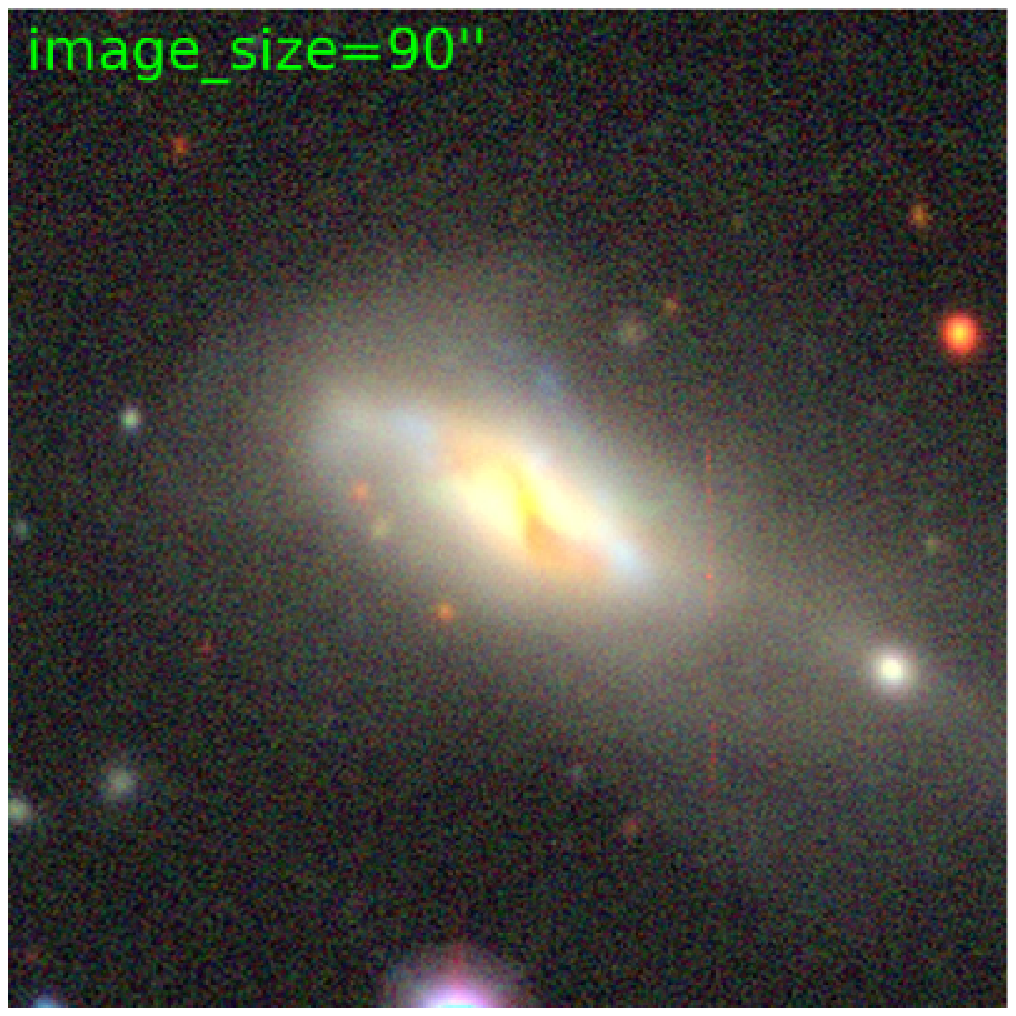}
\includegraphics[angle=0,width=0.174\textwidth]{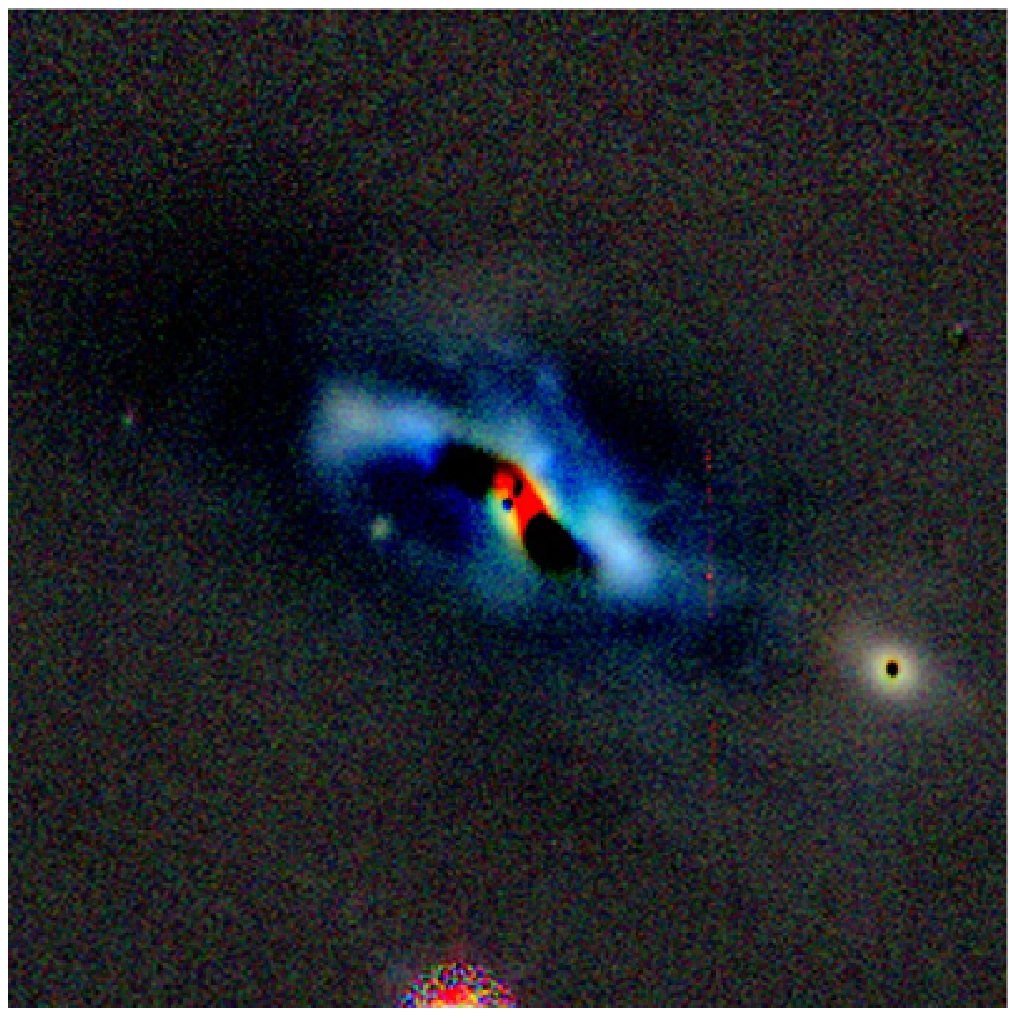}\\[0.5cm]
\includegraphics[angle=0,width=0.39\textwidth]{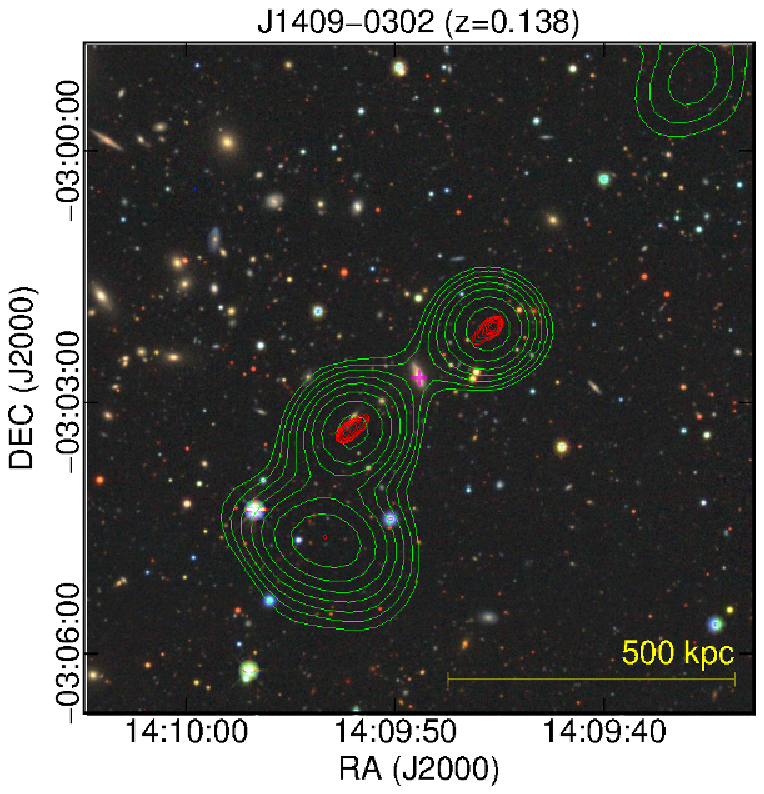}
\hspace{8mm}
\includegraphics[angle=0,width=0.39\textwidth]{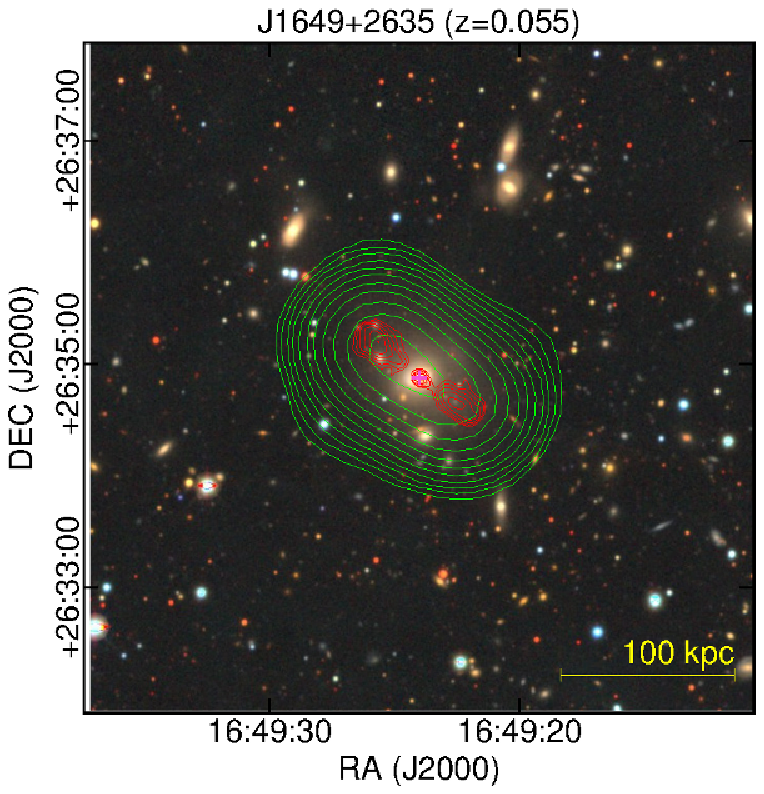}\\
\hspace{7mm}
\includegraphics[angle=0,width=0.174\textwidth]{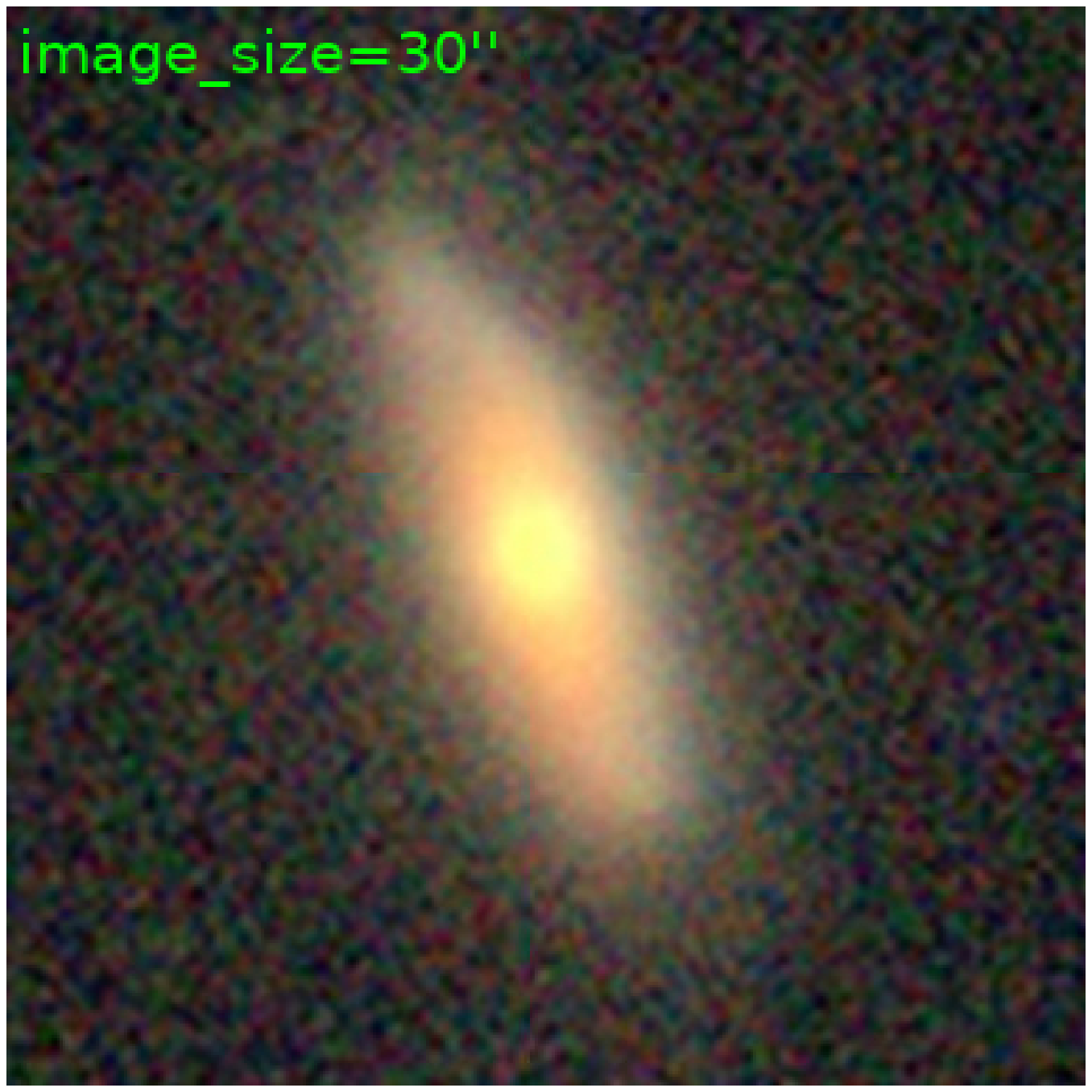}
\includegraphics[angle=0,width=0.174\textwidth]{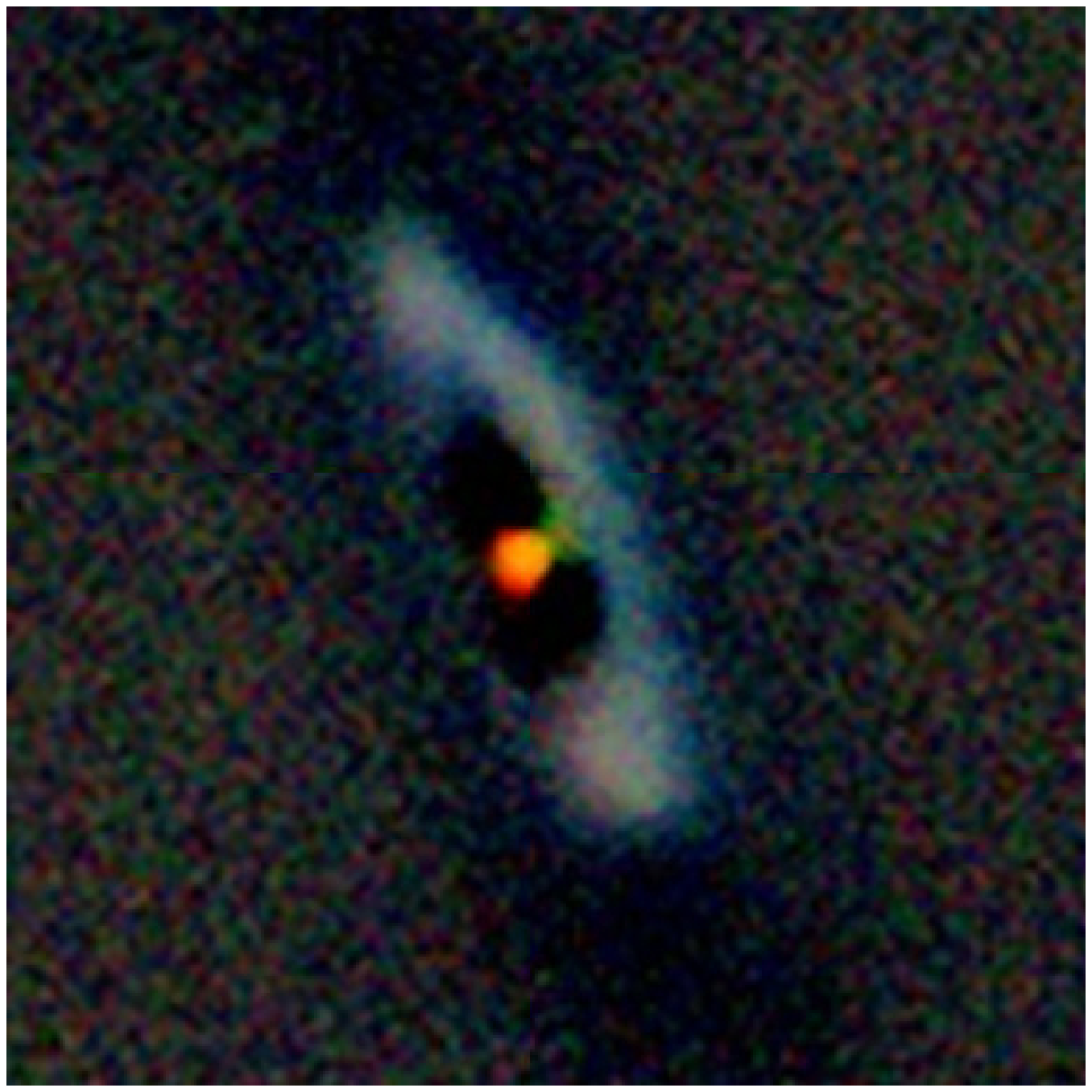}
\hspace{1.4cm}
\includegraphics[angle=0,width=0.174\textwidth]{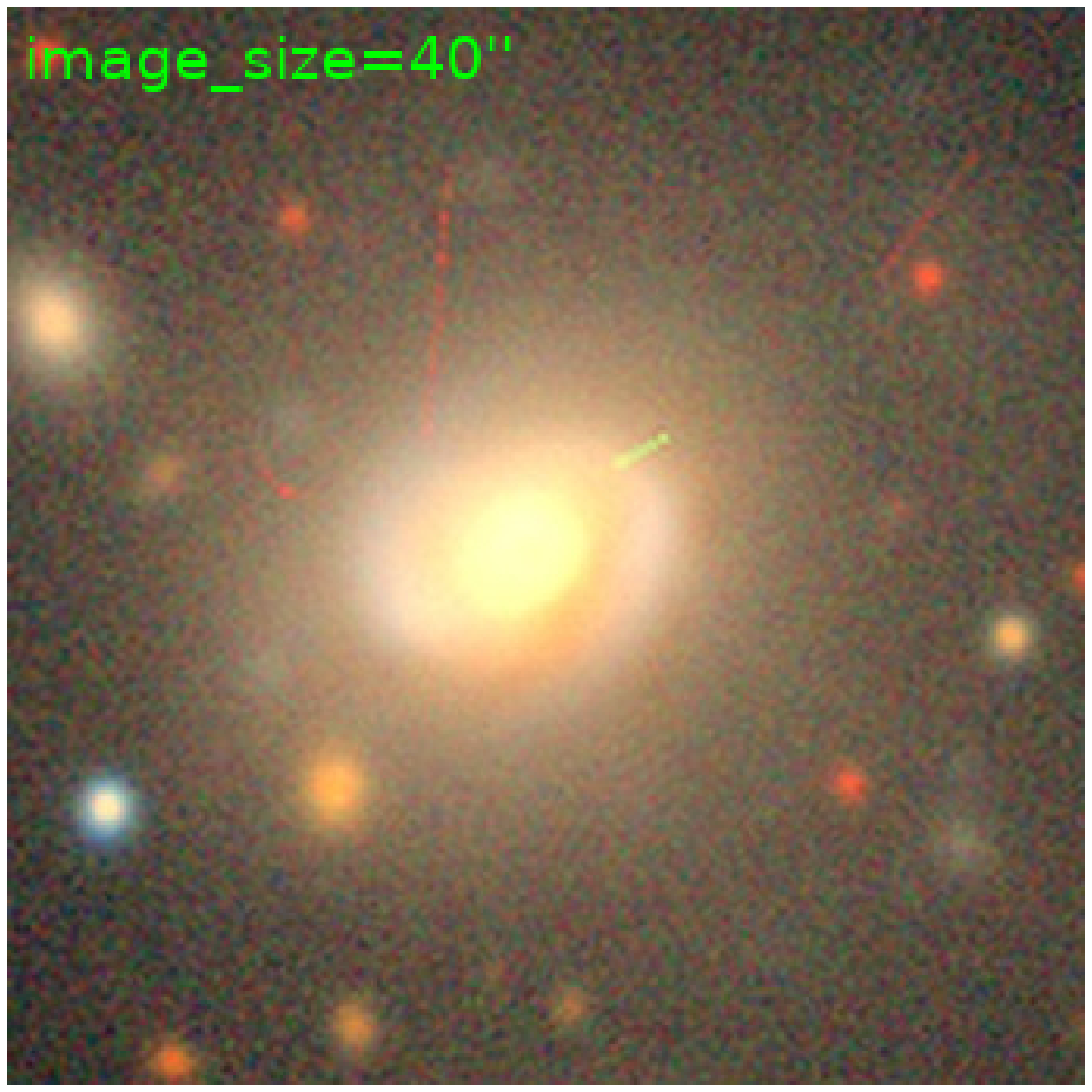}
\includegraphics[angle=0,width=0.174\textwidth]{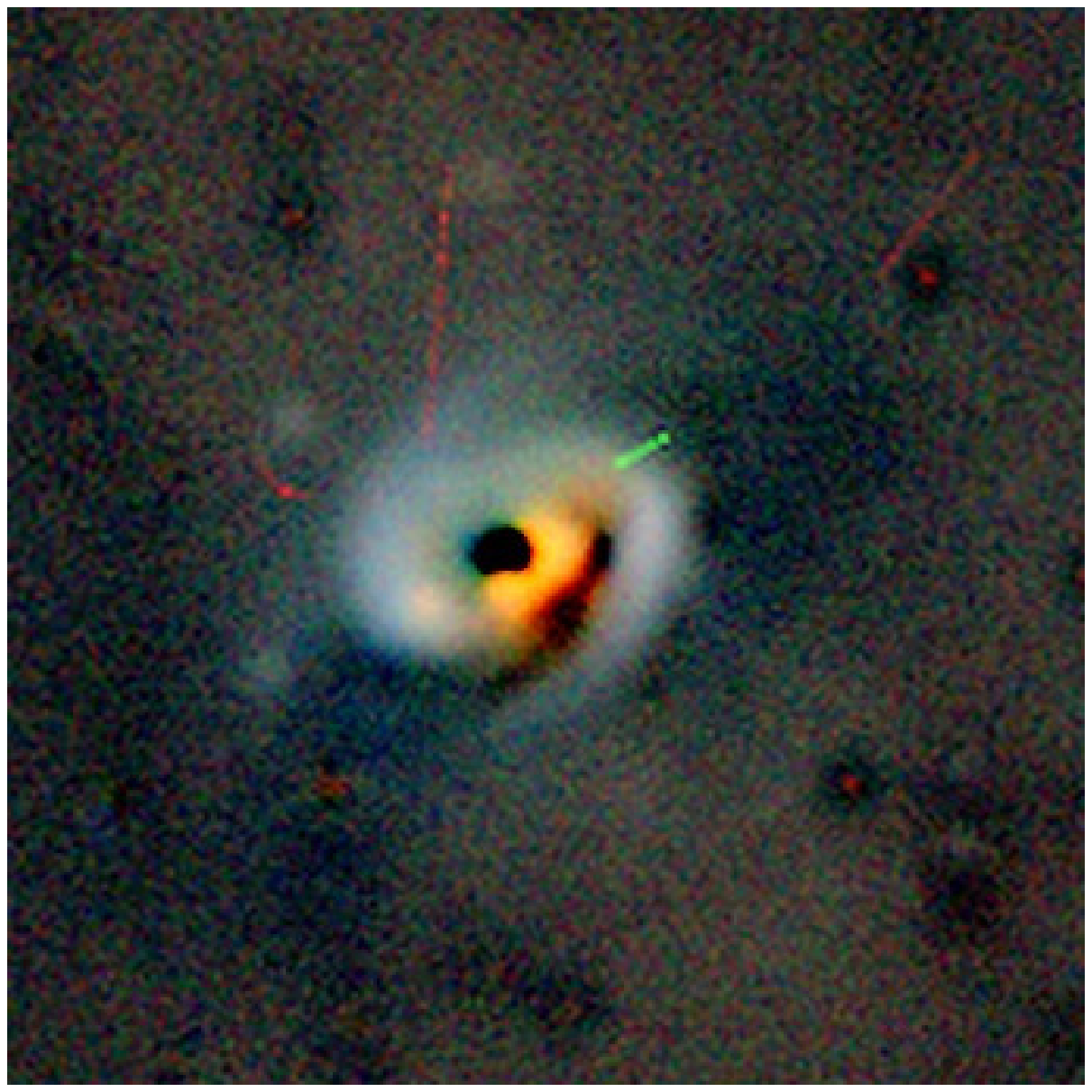}\\
\caption{ - {\it continued}}
\label{fig1}
\end{figure*}

By cross-matching the spiral sample taken from \citet{ks16} with the
radio catalogs of the NVSS \citep{ccg+98} and FIRST \citep{bwh95}, we
successfully identify eight double-lobed spiral galaxies, three of
which, J0326$-$0623, J1110+0321 and J1134+3046 are revealed for the
first time. J1128+2417 is another case that we independently
discovered in this work. However, it was recently reported by
\citet{Wu22} when we were preparing this manuscript for submission,
and we have to list it as a known case. We add a note for this object
in Sect.~\ref{Notes}. Another four previously known cases: J1159+5820
\citep{kjz+12}, J1352+3126 \citep{Donzelli07}, J1649+2635
\citep{mod+15} and the Speca \citep[J1409$-$0302,][]{hso+11} have also
been re-identified. Their recurrences validate our searching
strategy. By combining the radio and optical data, we show the
composite images for these eight spiral galaxies hosting double radio
lobes identified in this work in Fig.~\ref{fig1}.

For the other five known cases, J0836+0532 identified by
\citet{sis+15} is included in the catalog of \citet{ks16}, but without
morphological remarks and the classification certainty is $p$ = 0.23,
therefore it is missed. J0315$-$1906 \citep{lok98}, J0354$-$1340
\citep{Vietri22}, and J2318+4314 \citep{mmm+16} are not in the SDSS
sky, hence we cannot get them. J2345$-$0449 \citep{bvv+14} is not
included in \citet{ks16}. Among the 18 disk galaxies associated with
double radio lobes found in \citet{Wu22}, some of the central optical
galaxies seen face-on or having small inclination angles show
unambiguous spiral patterns, qualifying our selection based on
morphology. They belong to the same type of galaxies as discussed in
this work. Except for J1128+241, which is our target J1128+2417, we
examine the images by eye and pick out seven galaxies: J0209+0750,
J0219+0155, J0806+0624, J0832+1848, J1328+5710, J1656+6407 and
J1721+2624 as spirals subjectively. Since J0209+0750, J0806+0624,
J1328+5710 and J1656+6407 have also been claimed to show spiral
features by \citet{Wu22}, so that these four common targets are
included as the known cases of spirals in Table~\ref{sdragn}.

\subsection{Notes on the Newly Identified Spirals with Double Radio Lobes}
\label{Notes}
\subsubsection{J0326$-$0623}
J0326$-$0623 is a face-on galaxy at a redshift of $z = 0.18$ with two
major spiral arms, clearly shown in the zoomed DESI image and the
model-subtracted residual image in Fig.~\ref{fig1}. This galaxy is the
brightest cluster galaxy (BCG) of a galaxy cluster in the catalog of
\citet{Yang07}, which contains 13 bright member galaxies of $M^{\rm
  e}_{\rm r}\le-20.5$ mag. The total flux density detected by NVSS at
1.4~GHz is about 6~mJy. However, the upper lobe is superimposed by a
point-like radio source as detected by FIRST, which is associated with
J032624$-$062212, a foreground galaxy at $z \sim 0.16$. The morphology
of the radio lobes is slightly bent. It has a size of $\sim$430~kpc as
inferred by the NVSS 5$\sigma$ contour.

\subsubsection{J1110+0321}
J1110+0321 is a blue galaxy at $z = 0.03$. The optical observational
and residual images displayed in Fig.~\ref{fig1} indicate spiral-arm
structures. This galaxy belongs to a galaxy group \citep{Tempel12},
which contains seven members brighter than $-20.5$ mag. The NVSS image
features an elongated morphology with two lobes close to each other,
and FIRST detects bright sources at the peak in each lobe. The
inner-west component detected by FIRST is probably partially
associated with a background quasar QSO B1107+0337 at $z=0.965$. The
overall scale of the radio lobes measured based on the NVSS 5$\sigma$
contour is about 100~kpc.

\subsubsection{J1128+2417}
J1128+2417 is a blue galaxy at $z = 0.169$. The optical residual image
for this galaxy presents a faint imprint of spiral patterns, while the
high-quality deeper image from the Hubble Space Telescope \citep{Wu22}
clearly shows the existence of the spiral structures. With the method
introduced in Section~\ref{envir}, we find this galaxy is a satellite
galaxy in a galaxy group, which contains four members with $M^{\rm
  e}_{\rm r}\le-20.5$ mag. The NVSS map shows unresolved radio lobes
with an elongated morphology, but FIRST detects two bright jets with
some bridge emission. The scale for the radio emission indicated by
the NVSS is around 380~kpc.

\subsubsection{J1134+3046}
J1134+3046 is also a blue galaxy at $z = 0.046$. The optical residual
map of this galaxy presents clear structures of spiral arms. This
galaxy is a member in a galaxy group \citep{Tempel12}, which contains
four bright member galaxies with $M^{\rm e}_{\rm r}\le-20.5$
mag. Similar to J1110+0321 and J1128+2417, the NVSS map of J1134+3046
shows unresolved radio lobes with an elongated morphology, while the
FIRST image presents clear jets. The overall scale of radio emission
presented by the NVSS map is approximately 190~kpc.

\subsection{Relation between Radio Power and Stellar Mass of the Host Galaxy}

The double radio lobes of the spiral galaxies come from their central
SMBH. It is therefore natural to speculate that the power of these
radio lobes could be related to the mass of the SMBH. The mass of the
SMBH is difficult to assess directly. However, it is related to the
mass of the host galaxy \citep[e.g.,][]{fm00,tgb+02,mh03}.
\citet{Wu22} reported a positive correlation between $L_{\rm1.4~GHz}$
and the mass of SMBH $M_{BH}$ for nine previously known cases
with/without their 18 new disk galaxies hosting double radio lobes
(see their Figure 9). They estimated the stellar mass of the galaxy
$M_{*}$ by using SDSS multi-band photometry. Alternatively, the total
stellar mass of the host galaxy can be well estimated based on the
infrared luminosity of the galaxy which is less affected by star
formation history than the corresponding optical luminosity
\citep{bmk+03,wwz+13}. \citet{wh21} found a good scaling relation
between the stellar mass of the galaxy and the 3.4~$\mu$m luminosity
from the Wide-field Infrared Survey Explorer
\citep[WISE,][]{wem+10}. Following their procedure, we estimated the
stellar mass of each galaxy listed in Table~\ref{sdragn}.

\begin{figure}
\centering
\includegraphics[angle=0,width=0.45\textwidth]{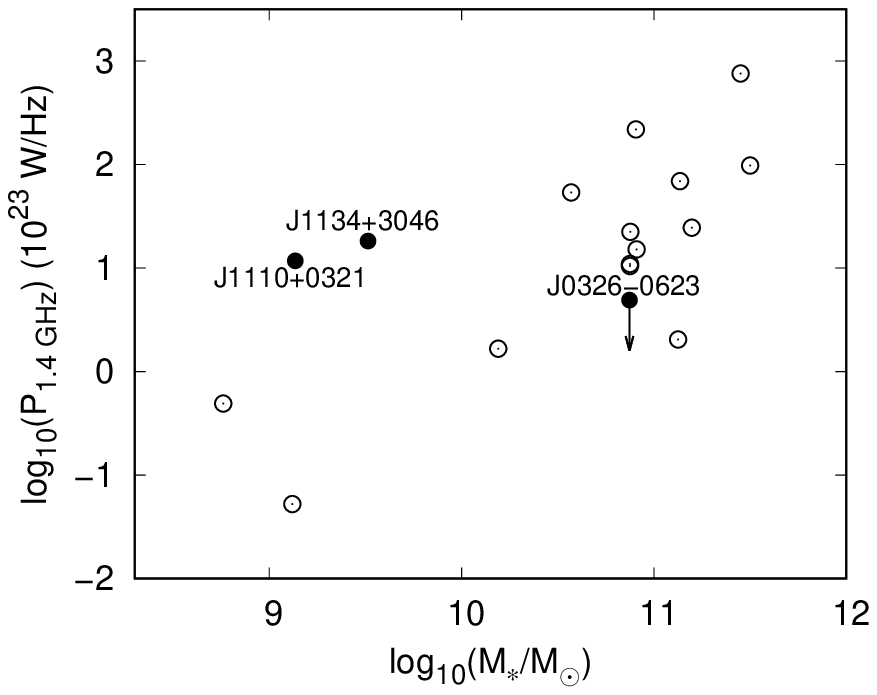}
\caption{Radio power versus stellar mass for 14 known (open) and three
  newly identified (solid, name labeled) spiral galaxies hosting
  double radio lobes.}
\label{magnitude}
\end{figure}

On the other hand, the 1.4~GHz flux densities for the radio lobes of
all these galaxies were obtained from the NVSS catalog, and the radio
powers were calculated via
\begin{equation}
P_{\rm 1.4~GHz}=4\pi D_{\rm L}^{2}\times S_{\rm 1.4~GHz}\times (1+z)^{1-\beta},
\label{power}
\end{equation}
where $P_{\rm 1.4~GHz}$ is the radio power in the unit of $10^{23}$
W~Hz$^{-1}$ and $D_{\rm L}=(1+z) \frac{c}{H_0} \int_{0}^{z}\frac{dz'}
{\sqrt{\Omega_{m}(1+z')^3+\Omega_{\Lambda}}}$ is the luminosity
distance of a galaxy at a redshift $z$. $S_{\rm 1.4~GHz}$ is the
1.4~GHz total flux density of the radio lobes in mJy extracted from
the NVSS catalog. $(1+z)^{(1-\beta)}$ is the {\it k}-correction term
and $\beta$ is the spectral index of radio galaxies. We adopted the
statistical mean of $\beta=0.74$ as obtained by \citet{lm07}.  All the
radio flux densities and the corresponding radio powers are listed in
Table~\ref{sdragn}. The radio powers are further compared with the
stellar mass $M_{*}$ of 14 known (open) and three newly identified
(solid) double-lobed spiral galaxies in Figure~\ref{magnitude}.

As shown in Figure~\ref{magnitude}, a positive correlation exists
between the stellar mass of spiral galaxy and the power of their
associated double radio lobes. We test the correlation by the Spearman
rank-order correlation, and get the significance $p_{s} = 0.00$,
indicating the existence of strong dependence between $M_{*}$ and
radio power of the lobes, with a relation of $P_{\rm 1.4~GHz} \propto
M_{*}^{0.8}$.

\subsection{Environment of the Spiral-hosted Double-lobed Sources}
\label{envir}

The mechanism for powering the large-scale double radio lobes by
spiral galaxies is not clear. Physically, radio lobes may be related
to a dense environment. \citet{hso+11} pointed out that J1409$-$0302
belongs to a galaxy cluster of MaxBCG J212.45357$-$03.04237 and it is
the central BCG. Its relic radio lobes may result from the accretion
of galactic filaments. Based on the morphology, \citet{sis+15}
suggested a merger scenario between a spiral and an elliptical galaxy
for both J1159+5820 and J1352+3126. They also noticed that J0836+0532
and J1352+3126 are in galaxy groups with very limited group members,
but listed them as field galaxies together with J1159+5820.
\citet{mod+15} found that J1649+2635 is in a group rather than a
cluster environment and may interact with another group, in which the
bright galaxy SDSS J164933.52+265052.0 resides.

With all such accumulated samples and the new discoveries as listed in
Table~\ref{sdragn}, we can have a good statistics on the environment
of these spiral galaxies. Based on the galaxy group/cluster catalogs
\citep[e.g.][]{Yang07, Tempel12, Tully15, Tempel18}, we find that most
of them are located in a galaxy group or a cluster. For J0354$-$1340,
the information is not available.

We further use the SDSS data to evaluate the richness of their parent
system. We follow the procedures of \citet{whl12} and \citet{wh15} by
counting the member galaxies with $M^{\rm e}_{\rm r}\le-20.5$ mag if
they have a velocity difference of 2500 km~s$^{-1}$ from the group or
cluster when the spectroscopic redshift is available or have a
redshift difference of $0.04(1+z)$ if only photometric redshifts are
available. Here, $M^{\rm e}_{\rm r}$ is evolution-corrected from
$M_{\rm r}$ with $M^{\rm e}_{\rm r}=M_{\rm r}+1.16z$. The number of
such bright member galaxies $N_{\rm gal}$ is listed in
Table~\ref{sdragn}. We notice that the member galaxies in the parent
system of these spirals with double radio lobes are much less than
those in the cluster catalog of \citet{whl12}. Based on the numbers of
member galaxies counted in the above way, we here call the system a
``cluster'' if ten or more members are included, or a ``group'' if
less than 10 members are found. In addition, we found that more than
half of these spirals with double radio lobes are the BCG or brightest
group galaxy (BGG) in the parent system.

We pick seven﻿ spiral galaxies from \citet{Wu22} and four of them are
listed in Table~\ref{sdragn} as known cases. We check the environment
for the rest three: J0219+0155, J0832+1848 and J1721+2624 following
the method described above. All of them are found in galaxy groups and
clusters \citep{Yang07, Tempel12, Tully15}, and are the BGGs
(J0219+0155: $N_{gal} = 6$; J0832+1848: $N_{gal} = 3$) and BCG
(J1721+2624: $N_{gal} = 15$). J1646+3831 is one of the six galaxies
claimed to show spiral structures in \citet{Wu22} besides J1128+2417
and the four cases. It is the BGG in a galaxy group ($N_{gal} = 5$)
listed in \citet{Tempel12}. All the evidence supports that the spirals
producing large-scale radio jets tend to reside in galaxy groups and
poor clusters.

\section{Concluding Remarks}
\label{sect:summary}
By cross-matching a large sample of machine-selected spiral galaxies
from the SDSS \citep{yaa+00} DR8 \citep{ks16} with the full radio
source catalogs of the NVSS \citep{ccg+98} and the FIRST
\citep{bwh95}, we identify three new spirals, J0326$–$0623, J1110+0321
and J1134+3046 hosting double radio lobes, together with five
previously known double-lobed spirals.

With the largest sample of double-lobed spiral galaxies by far, we
confirm that more massive spirals could produce more powerful
large-scale radio jets. We notice that most spiral galaxies that host
double radio lobes are located in galaxy groups or galaxy clusters and
more than a half of them are BGGs or BCGs, implying that the formation
of double radio lobes may be highly related to their surrounding
environment. A more noteworthy fact is that the galaxy groups or
clusters in which these spirals reside have very limited members,
i.e. the environmental density is denser than the field, but not so
dense and hot as in the center of rich clusters where spirals may be
destroyed.

\begin{acknowledgements}
We thank the anonymous referee for helpful comments. The authors are
supported by the National Natural Science Foundation of China (NSFC,
Grant Nos. 11988101, 11833009), the National SKA Program of China
(grant No. 2022SKA0120103), the National Key R\&D Program of China
(Nos. 2021YFA1600401 and 2021YFA1600400), and the Open Project Program
of the Key Laboratory of FAST, NAOC, Chinese Academy of
Sciences. X.Y.G acknowledges the financial support from the CAS-NWO
cooperation programme (grant No. GJHZ1865).
The National Radio Astronomy Observatory is a facility of the National
Science Foundation operated under cooperative agreement by Associated
Universities, Inc.
Funding for the Sloan Digital Sky Survey IV has been provided by the
Alfred P. Sloan Foundation, the U.S. Department of Energy Office of
Science, and the Participating Institutions. SDSS acknowledges support
and resources from the Center for High-Performance Computing at the
University of Utah. The SDSS website is www.sdss.org.
\end{acknowledgements}

\bibliographystyle{raa}
\bibliography{ref}

\begin{thebibliography}{40}
\providecommand\natexlab[1]{#1}
\providecommand\JournalTitle[1]{#1}

\bibitem[{Bagchi} {et~al.}(2014)]{bvv+14}
{Bagchi}, J., {Vivek}, M., {Vikram}, V., {et~al.} 2014, \apj, 788, 174

\bibitem[{Becker} {et~al.}(1995)]{bwh95}
{Becker}, R.~H., {White}, R.~L., \& {Helfand}, D.~J. 1995, \apj, 450, 559

\bibitem[{Bell} {et~al.}(2003)]{bmk+03}
{Bell}, E.~F., {McIntosh}, D.~H., {Katz}, N., \& {Weinberg}, M.~D. 2003, \apjs,
  149, 289

\bibitem[{Chen} {et~al.}(2020)]{Chen20}
{Chen}, S., {J{\"a}rvel{\"a}}, E., {Crepaldi}, L., {et~al.} 2020, \mnras, 498,
  1278

\bibitem[{Condon} {et~al.}(1998)]{ccg+98}
{Condon}, J.~J., {Cotton}, W.~D., {Greisen}, E.~W., {et~al.} 1998, \aj, 115,
  1693

\bibitem[{Dey} {et~al.}(2019)]{dsl+19}
{Dey}, A., {Schlegel}, D.~J., {Lang}, D., {et~al.} 2019, \aj, 157, 168

\bibitem[{Donzelli} {et~al.}(2007)]{Donzelli07}
{Donzelli}, C.~J., {Chiaberge}, M., {Macchetto}, F.~D., {et~al.} 2007, \apj,
  667, 780

\bibitem[{Ferrarese} \& {Merritt}(2000)]{fm00}
{Ferrarese}, L., \& {Merritt}, D. 2000, \apjl, 539, L9

\bibitem[{Hota} {et~al.}(2011)]{hso+11}
{Hota}, A., {Sirothia}, S.~K., {Ohyama}, Y., {et~al.} 2011, \mnras, 417, L36

\bibitem[{Huertas-Company} {et~al.}(2011)]{Huertas11}
{Huertas-Company}, M., {Aguerri}, J.~A.~L., {Bernardi}, M., {Mei}, S., \&
  {S{\'a}nchez Almeida}, J. 2011, \aap, 525, A157

\bibitem[{Keel} {et~al.}(2006)]{kwol06}
{Keel}, W.~C., {White}, Raymond~E., I., {Owen}, F.~N., \& {Ledlow}, M.~J. 2006,
  \aj, 132, 2233

\bibitem[{Kimball} \& {Ivezi{\'c}}(2008)]{Kimball08}
{Kimball}, A.~E., \& {Ivezi{\'c}}, {\v Z}. 2008, \aj, 136, 684

\bibitem[{Kozie{\l}-Wierzbowska} {et~al.}(2012)]{kjz+12}
{Kozie{\l}-Wierzbowska}, D., {Jamrozy}, M., {Zola}, S., {Stachowski}, G., \&
  {Ku{\'z}micz}, A. 2012, \mnras, 422, 1546

\bibitem[{Kuminski} \& {Shamir}(2016)]{ks16}
{Kuminski}, E., \& {Shamir}, L. 2016, \apjs, 223, 20

\bibitem[{Ledlow} {et~al.}(1998)]{lok98}
{Ledlow}, M.~J., {Owen}, F.~N., \& {Keel}, W.~C. 1998, \apj, 495, 227

\bibitem[{Ledlow} {et~al.}(2001)]{loyh01}
{Ledlow}, M.~J., {Owen}, F.~N., {Yun}, M.~S., \& {Hill}, J.~M. 2001, \apj, 552,
  120

\bibitem[{Lin} \& {Mohr}(2007)]{lm07}
{Lin}, Y.-T., \& {Mohr}, J.~J. 2007, \apjs, 170, 71

\bibitem[{Lintott} {et~al.}(2008)]{lss+08}
{Lintott}, C.~J., {Schawinski}, K., {Slosar}, A., {et~al.} 2008, \mnras, 389,
  1179

\bibitem[{Mao} {et~al.}(2015)]{mod+15}
{Mao}, M.~Y., {Owen}, F., {Duffin}, R., {et~al.} 2015, \mnras, 446, 4176

\bibitem[{Marconi} \& {Hunt}(2003)]{mh03}
{Marconi}, A., \& {Hunt}, L.~K. 2003, \apjl, 589, L21

\bibitem[{Meert} {et~al.}(2015)]{mvb15}
{Meert}, A., {Vikram}, V., \& {Bernardi}, M. 2015, \mnras, 446, 3943

\bibitem[{Mulcahy} {et~al.}(2016)]{mmm+16}
{Mulcahy}, D.~D., {Mao}, M.~Y., {Mitsuishi}, I., {et~al.} 2016, \aap, 595, L8

\bibitem[{Ortiz Mart{\'\i}nez} \& {Andernach}(2016)]{Martinez16}
{Ortiz Mart{\'\i}nez}, A.~F., \& {Andernach}, H. 2016, arXiv e-prints,
  arXiv:1610.02572

\bibitem[{Saulder} {et~al.}(2016)]{svc+16}
{Saulder}, C., {van Kampen}, E., {Chilingarian}, I.~V., {Mieske}, S., \&
  {Zeilinger}, W.~W. 2016, \aap, 596, A14

\bibitem[{Singh} {et~al.}(2015)]{sis+15}
{Singh}, V., {Ishwara-Chandra}, C.~H., {Sievers}, J., {et~al.} 2015, \mnras,
  454, 1556

\bibitem[{Tempel} {et~al.}(2018)]{Tempel18}
{Tempel}, E., {Kruuse}, M., {Kipper}, R., {et~al.} 2018, \aap, 618, A81

\bibitem[{Tempel} {et~al.}(2012)]{Tempel12}
{Tempel}, E., {Tago}, E., \& {Liivam{\"a}gi}, L.~J. 2012, \aap, 540, A106

\bibitem[{Tremaine} {et~al.}(2002)]{tgb+02}
{Tremaine}, S., {Gebhardt}, K., {Bender}, R., {et~al.} 2002, \apj, 574, 740

\bibitem[{Tully}(2015)]{Tully15}
{Tully}, R.~B. 2015, \aj, 149, 171

\bibitem[{Vietri} {et~al.}(2022)]{Vietri22}
{Vietri}, A., {J{\"a}rvel{\"a}}, E., {Berton}, M., {et~al.} 2022, \aap, 662,
  A20

\bibitem[{Wen} {et~al.}(2013)]{wwz+13}
{Wen}, X.-Q., {Wu}, H., {Zhu}, Y.-N., {et~al.} 2013, \mnras, 433, 2946

\bibitem[{Wen} \& {Han}(2015)]{wh15}
{Wen}, Z.~L., \& {Han}, J.~L. 2015, \apj, 807, 178

\bibitem[{Wen} \& {Han}(2021)]{wh21}
{Wen}, Z.~L., \& {Han}, J.~L. 2021, \mnras, 500, 1003

\bibitem[{Wen} {et~al.}(2012)]{whl12}
{Wen}, Z.~L., {Han}, J.~L., \& {Liu}, F.~S. 2012, \apjs, 199, 34

\bibitem[{Willett} {et~al.}(2013)]{Willett13}
{Willett}, K.~W., {Lintott}, C.~J., {Bamford}, S.~P., {et~al.} 2013, \mnras,
  435, 2835

\bibitem[{Wright} {et~al.}(2010)]{wem+10}
{Wright}, E.~L., {Eisenhardt}, P. R.~M., {Mainzer}, A.~K., {et~al.} 2010, \aj,
  140, 1868

\bibitem[{Wu} {et~al.}(2022)]{Wu22}
{Wu}, Z., {Ho}, L.~C., \& {Zhuang}, M.-Y. 2022, \apj, 941, 95

\bibitem[{Yang} {et~al.}(2007)]{Yang07}
{Yang}, X., {Mo}, H.~J., {van den Bosch}, F.~C., {et~al.} 2007, \apj, 671, 153

\bibitem[{York} {et~al.}(2000)]{yaa+00}
{York}, D.~G., {Adelman}, J., {Anderson}, John~E., J., {et~al.} 2000, \aj, 120,
  1579

\bibitem[{Yuan} {et~al.}(2016)]{yhw16}
{Yuan}, Z.~S., {Han}, J.~L., \& {Wen}, Z.~L. 2016, \mnras, 460, 3669

\end{thebibliography}

\end{document}